\documentstyle[12pt,subeqnarray,psfig]{article}
\begin{document}
%
%
%
%
\newenvironment{lefteqnarray}{\arraycolsep=0pt\begin{eqnarray}}
{\end{eqnarray}\protect\aftergroup\ignorespaces}
\newenvironment{lefteqnarray*}{\arraycolsep=0pt\begin{eqnarray*}}
{\end{eqnarray*}\protect\aftergroup\ignorespaces}
\newenvironment{leftsubeqnarray}{\arraycolsep=0pt\begin{subeqnarray}}
{\end{subeqnarray}\protect\aftergroup\ignorespaces}
\newcommand{\displayfrac}[2]{\frac{\displaystyle #1}{\displaystyle #2}}
\newcommand{\diff}{{\rm\,d}}
\newcommand{\img}{{\rm i}}
\newcommand{\appleq}{\stackrel{<}{\sim}}
\newcommand{\appgeq}{\stackrel{>}{\sim}}
\newcommand{\Int}{\mathop{\rm Int}\nolimits}
\newcommand{\Nint}{\mathop{\rm Nint}\nolimits}
\newcommand{\Min}{\mathop{\rm min}\nolimits}
\newcommand{\arcsinh}{\mathop{\rm arcsinh}\nolimits}

\title{Homeoidally striated density profiles. II. \\
       Anisotropy and rotation}


\author{{R. Caimmi}\footnote{
{\it Astronomy Department, Padua Univ., Vicolo Osservatorio 2,
I-35122 Padova, Italy}
email: caimmi@pd.astro.it}
\phantom{agga}}


%
\maketitle
\begin{quotation}
\section*{}
\begin{Large}
\begin{center}

Abstract

\end{center}
\end{Large}
\begin{small}


With regard to homeoidally striated
Jacobi ellipsoids (Caimmi \& Marmo 2005), a
unified theory of systematically rotating and
peculiar motions is developed, where both real
and imaginary rotation (around a fixed axis)
are considered.   The effect of positive or
negative residual motion excess, is shown
to be equivalent to an additional real or
imaginary rotation, respectively.   Then it
is realized that a homeoidally striated Jacobi
ellipsoid with assigned velocity distribution,
always admits an adjoint configuration i.e. a
classical Jacobi ellipsoid of equal mass and
axes.  In addition, further constraints are
established on the amount of residual velocity
anisotropy along the principal axes, for
triaxial configurations.   Special effort is
devoted to investigating sequences of virial
equilibrium configurations in terms of
normalized parameters, including the effects
of both density and velocity profile.   In
particular, it is shown that bifurcation
points from axisymmetric to triaxial 
configurations occur as in classical Jacobi
ellipsoids, contrary to earlier results
(Wiegandt 1982a,b; Caimmi \& Marmo 2005).
The reasons of the above mentioned discrepancy
are also explained.   The occurrence of
centrifugal support at the ends of major equatorial
semiaxis, is briefly discussed.   An
interpretation of recent results from
numerical simulations on stability (Meza
2002), is provided in the light of the model.
A physical interpretation of the early Hubble
sequence is shortly reviewed and discussed
from the standpoint of the model.   More
specifically, (i) elliptical galaxies are
considered as isolated systems,
and an allowed region within
Ellipsoidland (Hunter \& de Zeeuw 1997),
related to the occurrence of bifurcation
points from ellipsoidal to pear-shaped
configurations, is shown to be consistent
with observations; (ii) elliptical galaxies
are considered as embedded within dark
matter haloes and, under reasonable
assumptions, it is shown that tidal effects
from embedding haloes have little influence
on the above mentioned results; (iii) dark
matter haloes and hosted elliptical galaxies,
idealized as a single homeoidally striated
Jacobi ellipsoid, are considered in connection
with the cosmological transition from expansion
to relaxation, by generalizing an earlier model
(Thuan \& Gott 1975), and the existence of a
lower limit to the flattening of relaxed
(oblate-like) configurations, is established.
On the other hand,  no lower limit is found
to the elongation of relaxed (prolate-like)
configurations, and the
observed lack of elliptical galaxies more
elongated than $E7$ needs a
different physical interpretation, such as
the occurrence of bending instabilities
(Polyachenko \&
Shukhman 1979; Merritt \& Hernquist 1991).

\noindent
{\it keywords - cosmology: dark matter - galaxies:
evolution - galaxies: formation - galaxies: haloes -
galaxies: structure.}

\end{small}
\end{quotation}
%

\section{Introduction}\label{intro}
Large-scale celestial objects, represented
as self-gravitating fluids, exhibit different
features according if their subunits, or
``particles'', are ``collisional'' or
``collisionless''.

In the former alternative, the gravitational
field that is generated by the system as a
whole is negligible in respect of the force
between two colliding subunits, when they 
repel each other strongly.   Consequently,
particles in self-gravitating, collisional
fluids, are subjected to violent and 
short-lived accelerations as they are 
sufficiently close to each other, interspersed
with longer periods when they move at nearly 
constant velocity.

In the latter alternative, the gravitational 
field that is generated by the system as a whole
is dominant in respect of the force between two
subunits, even if they are close to each other.
Consequently, particles in self-gravitating,
collisionless fluids, are subjected to smooth
and long-lived accelerations through their
trajectories.

Gas in stars and stars in stellar systems may
be approximated, to a good extent, as collisional
and collisionless, ideal self-gravitating fluids,
respectively.   The statistical problem of 
particle motion in the latter case is that
of the former, with the collisions left out.
``Ideal'' has to be intended
as ``particles collide as perfectly and undeformable
spheres'' in the former situation, and ``particles
do not interact each other at all'' in the latter.

The motion equation of fluid flow turns out to be 
the same for both collisional and collisionless fluids
(e.g., Jeans 1929, Chap.\,II, \S\,26, Chap.\,VII, 
\S\S\,211-215; Binney \& Tremaine 1987, Chap.\,4, 
\S\,4.2) provided gases are thought of as far from
equilibrium. Then the velocity distribution of
molecules no longer obeys Maxwell distribution
and is, in general, anisotropic; accordingly, the
pressure is represented by a stress tensor.   In
the special case of isotropic velocity distributions,
the pressure attains its usual meaning and the
motion equation reduces to Euler's equation,
provided the velocity of an infinitesimal fluid
element is intended as the mean velocity of all
the particles within the same element at the time
considered.   On the contrary, Euler's equation
for a given fluid with isotropic distribution
can be generalized to an anisotropic one, by
replacing velocities with mean velocities and
pressures with stress tensors, in the sense
specified above. 

For special classes of ideal, self-gravitating 
fluids, such as steadily rotating polytropes
with polytropic index $n\ge1/2$ (Vandervoort
1980a) and isothermal spheres (e.g., Binney \&
Tremaine 1987, Chap.\,4, \S\,4.4b), a one-to-one
correspondence has been discovered between
collisional and collisionless systems with
equal physical parameters.   Steadily rotating
polytropes may be represented, to a first extent,
as steadily rotating, homeoidally striated 
ellipsoids (e.g., Vandervoort 1980b; Vandervoort 
\& Welty 1981; Lai et al. 1993).

Though most astronomical bodies exhibit
ellipsoidal-like shapes and isopycnic
surfaces, still some caution must be used
in dealing with the above approximation,
with regard to local values of physical
parameters.   This is why steadily
rotating polytropes are in hydrostatic 
equilibrium while, owing to Hamy's theorem
(e.g., Chambat 1994), the contrary holds
for their homeoidally striated ellipsoidal
counterparts.   On the other hand, homeoidally
striated ellipsoids may safely be assumed
as a viable approximation to self-gravitating
fluids, with regard to typical values of
physical parameters, averaged over the whole
volume
\footnote{An important exception is the energy    
ratio of rotational to random motions, $E_       
{rot}/E_{pec}$ (Caimmi 1979, 1983).}%
(e.g., Vandervoort 1980b; Vandervoort \& 
Welty 1981; Lai et al. 1993).

Celestial objects have been modelled as                                                     
collisional, self-gravitating fluids,
in particular homeoidally striated
ellipsoids, since the beginning of the
scientific era (see e.g., Chandrasekhar
1969, Chap.\,1).  The evidence of rotation
in spiral galaxies and the symmetry of
figure shown by elliptical galaxies and
spiral bulges, suggested the following
(e.g., Jeans 1929, Chap.\,XIII, \S\,299):
(i) all (regular) galaxies rotate; (ii)
the symmetry of figure is precisely such
as rotation might be expected to produce;
(iii) the observed shapes of (regular)
galaxies can be explained as the figures
assumed by masses rotating under their
own gravitation.  In particular, the system
of primeval stars must have conserved
roughly the same shape as the original mass
of gas from which it emerged (e.g., Blaauw
1965).

The above outlined ideas were accepted up
to about thirthy years ago, when observations
yielded increasing evidence, that giant 
elliptical galaxies cannot be sustained by
systematic
rotation (e.g., Bertola \& Capaccioli 1975;
Binney 1976; Illingworth 1977, 1981; Schechter
\& Gunn 1979).   Then giant elliptical galaxies
must necessarily be conceived as collisionless,
self-gravitating fluids, owing their boundaries
to anisotropic peculiar velocity distribution
of stars (Binney 1976, 1978), with a negligible
contribution from figure rotation.   Accordingly,
the shape corresponds - in general - to triaxial
\footnote{Strictly speaking, all tridimensional
bodies may be conceived as triaxial, in particular
spheres and spheroids.   Throughout this paper,
``triaxial'' shall be intended as denoting
ellipsoids where the axes are different in length.}%
ellipsoids.   Owing to high-resolution simulations,
the same holds also for (nonbaryonic) dark matter
haloes hosting galaxies and cluster of galaxies
(e.g., Hoeft et al. 2004; Rasia et al. 2004;
Bailin \& Steinmetz 2004).

Isotropic peculiar velocity distributions
necessarily imply configurations which
rotate around the minor axis.   Accordingly,
empirical evidence of systematic rotation
around the major axis (e.g., Bertola \&
Galletta 1978), in absence of tidal potential,
makes a signature of the occurrence of
anisotropic peculiar velocity distribution.

Though anisotropy in peculiar velocity distribution
is a basic ingredient in the description and 
investigation of stellar systems and hosting
dark matter haloes, conceived as self-gravitating
collisionless fluids (e.g., Binney 1976, 1978, 1980),
still no attempt (to the knowledge of the author)
has been made to gain more insight into the
intrinsic connection between systematic
and random motions, and the extent to which they
are related in determining the figure of equilibrium.
The current paper aims to overhelm the above mentioned
gap, using a theory where systematic and random motions
are unified.

To this aim, the effect of anisotropy in peculiar
velocity distribution shall be related to the effect
of systematic (real or imaginary) rotation (Caimmi
1996b).   It will be shown the existence of a
one-to-one correspondance between configurations 
characterized by systematic ({\it real}) rotation
and {\it anisotropic} peculiar velocity distribution,
on one hand, and configurations characterized by
systematic ({\it real} or {\it imaginary}; 
{\it prograde} and/or {\it retrograde}) rotation and
{\it isotropic} peculiar velocity distribution.

An alternative explanation for
the absence of elliptical galaxies
(and nonbaryonic dark matter haloes)
more flattened or elongated than
$E7$ as due to bending
instabilities, has been suggested long
time ago from analytical considerations
involving homogeneous (oblate and
prolate) spheroids
(Polyachenko \& Shukhman 1979;
Fridman \& Polyachenko 1984, Vol.\,1,
Chap.\,4, Sect.\,3.3, see also
pp.\,313-322; Vol.\,2,
p.\,159) and numerical simulations
involving inhomogeneous (oblate and
prolate) spheroids
(Merritt \& Hernquist 1991; Merritt
\& Sellwood 1994).

In a cosmological
scenario (Thuan \& Gott 1975), the
occurrence of a limiting ellipticity
in oblate configurations depends on
the amount of spin growth regardless
from the onset of bending instabilities.
An interesting question could be if
a similar conclusion holds for prolate
configurations.   It will be seen
that the onset of bending instabilities
is necessary for the occurrence of a
limiting ellipticity in prolate
configurations.

The current paper is organized as follows.   The general
theory of homeoidally striated density profiles, and the
results of interest to the aim of this investigation,
are outlined in Sect.\,\ref{gente}.   A unified theory
of systematic and random motions, implying a definition
of imaginary rotation, is developed in Sect.\,\ref{unte} 
where, in addition, an interpretation is provided
to recent results from numerical simulations on 
stability.  An application of the above mentioned 
theory, namely a physical interpretation of the
early Hubble sequence, is performed in Sect.\,\ref{elli}, 
on general grounds, by considering (i) elliptical
galaxies as isolated systems; (ii) elliptical galaxies 
as embedded in dark matter haloes; (iii) dark
matter haloes and hosted elliptical galaxies,
idealized as a single homeoidally striated
Jacobi ellipsoid, in connection
with the cosmological transition from expansion to
relaxation; and establishing the existence of a
lower limit to the flattening of relaxed
(oblate-like) configurations.   Some
concluding remarks are drawn in Sect.\,\ref{conc},   
and a few arguments are treated with more detail in 
the Appendix.   

\section{General theory}
\label{gente}
\subsection{Homeoidally striated density profiles}
\label{profi}

A general theory for homeoidally striated density 
profiles has been developed in earlier approaches 
(Roberts 1962; Caimmi 1993a; Caimmi \& Marmo 2003, 
2005
\footnote{A more extended version is available at
the arxiv electronic site, as astro-ph/505306.},
hereafter quoted as CM03, CM05, respectively),
and an interested reader is addressed
therein for deeper insight.   What is relevant for
the current investigation, shall be mentioned and
further developed here.

The isopycnic (i.e. constant density) surfaces are 
defined by the following law:
\begin{leftsubeqnarray}
\slabel{eq:profga}
&& \rho=\rho_0f(\xi)~~;\quad f(1)=1~~;\quad\rho_0=
\rho(1)~~; \\
\slabel{eq:profgb}
&& \xi=\frac r{r_0}~~;\quad0\le\xi\le\Xi~~;\quad
\Xi=\frac R{r_0}~~;
\label{seq:profg}
\end{leftsubeqnarray}
where $\rho_0$, $r_0$, are a scaling density and
a scaling radius, respectively, related to a
reference isopycnic surface, $\xi$ is a scaled
distance, independent of the direction along
which radial coordinates are calculated, and
$\Xi=\xi(r)$, is related to the boundary, where
$r=R$.   Both cored
and cuspy density profiles, according to the 
explicit expression chosen for the scaled
density, $f(\xi)$, are represented by
Eqs.\,(\ref{seq:profg}).   In the former 
alternative, it is used a different normalization 
with respect to Caimmi (1993a), where $\xi=r/R$
and $\rho_0$ is the central density. 

The mass, the inertia tensor, and the potential
self-energy tensor are:
\begin{lefteqnarray}
\label{eq:M}
&& M=\nu_{mas}M_0~~; \\
\label{eq:Ipq}
&& I_{pq}=\delta_{pq}\nu_{inr}Ma_p^2~~; \\
\label{eq:Espq}
&& (E_{sel})_{pq}=-\frac{GM^2}{a_1}\nu_{sel}
(B_{sel})_{pq}=-\frac{GM^2}{a_1}{\cal S}_
{pq}~~; \\
\label{eq:Es}
&& E_{sel}=\sum_{i=1}^3(E_{sel})_{ii}=\frac
{GM^2}{a_1}\nu_{sel}
B_{sel}=-\frac{GM^2}{a_1}{\cal S}~~; \\
\label{eq:Bspq}
&& (B_{sel})_{pq}=\delta_{pq}\epsilon_{p2}
\epsilon_{p3}A_p~~;\quad B_{sel}=\sum_{s=1}^3
\epsilon_{s2}\epsilon_{s3}A_s~~; \\
\label{eq:Spq}
&& {\cal S}_{pq}=\nu_{sel}(B_{sel})_{pq}~~;
\quad{\cal S}=\nu_{sel}B_{sel}~~;
\end{lefteqnarray}
where $\delta_{pq}$ is the Kronecker symbol;
$G$ is the constant of gravitation; $\nu_
{mas}$, $\nu_{inr}$, $\nu_{sel}$, are profile
factors i.e. depend only on the density profile
via the scaled radius, $\Xi$; $a_1$, $a_2$,
$a_3$, are semiaxes; $\epsilon_{pq}=a_p/a_q$ 
are axis
ratios; $A_1$, $A_2$, $A_3$, are shape factors
i.e. depend only on the axis ratios; and $M_0$
is the mass of a homogeneous ellipsoid with
same density and boundary as the reference
isopycnic surface:
\begin{equation}
\label{eq:M0}
M_0=\frac{4\pi}3\rho_0a_{01}a_{02}a_{03}~~;
\end{equation}
where $a_{01}$, $a_{02}$, $a_{03}$, are the
semiaxes of the ellipsoid bounded by the 
reference isopycnic surface.

The limiting case of homogeneous configurations
reads:
\begin{leftsubeqnarray}
\slabel{eq:omoga}
&& f(\xi)=1~~,\quad0\le\xi\le\Xi~~; \\
\slabel{eq:omogb}
&& \nu_{mas}=\Xi^3~~;\quad\nu_{inr}=\frac15~~;
\quad\nu_{sel}=\frac3{10}~~;
\label{seq:omog}
\end{leftsubeqnarray}
for further details, see CM03, CM05.

\subsection{Homeoidally striated Jacobi ellipsoids}
\label{Jael}

In dealing with angular momentum and rotational
energy, the preservation of (triaxial) ellipsoidal
shape imposes severe constraints on the rotational
velocity field.    Leaving an exhaustive investigation
to more refined approaches, our attention shall
be limited here to the special case of homeoidally
striated Jacobi ellipsoids (CM05).   Accordingly,
the systematic velocity field is defined by the
law:
\begin{equation}
\label{eq:rvrot}
\frac{v_{rot}(r,\theta)}{v_{rot}(R,\theta)}=\frac
{v_{rot}(a_p^\prime,0)}{v_{rot}(a_p,0)}~~;
\end{equation}
or equivalently, the angular (with respect to
$x_3$ axis) velocity field is defined by the law:
\begin{equation}
\label{eq:rvang}
\frac{\Omega(r,\theta)}{\Omega(R,\theta)}=\frac
{\Omega(a_p^\prime,0)}{\Omega(a_p,0)}~~;
\end{equation}
where $(r,\theta)$, $(R,\theta)$, represent a
point on a generic isopycnic surface and on the
boundary, respectively, along a fixed radial
direction, and $(a_p^\prime,0)$, $(a_p,0)$,
$p=1,2,$ represent
the end of the corresponding equatorial semiaxis.

It is worth noticing that the following special
cases are described by Eqs.\,(\ref{eq:rvrot}) or
(\ref{eq:rvang}): (a) rigid rotation; (b) constant
rotation velocity everywhere; (c) rigid rotation
of isopycnic surfaces and constant rotation
velocity on the equatorial plane.   To maintain
ellipsoidal shapes, differential rotation [e.g.,
cases (b) and (c) above] must necessarily be
restricted to axysimmetric figures i.e. spheroidal
configurations.   In the limiting situation of
homogeneous, rigidly rotating, dynamical (or
hydrostatic) equilibrium configurations,
homeoidally striated Jacobi ellipsoids reduce
to classical Jacobi ellipsoids (CM05).

The angular-momentum vector and the rotational-energy
tensor are:
\begin{lefteqnarray}
\label{eq:Jpq}
&& J_s=\delta_{s3}\eta_{anm}\nu_{anm}M
a_p(1+\epsilon_{qp}^2)(v_{rot})_p~~;
\quad p\ne q\ne s~~; \\
\label{eq:Erpq}
&& (E_{rot})_{pq}=\delta_{pq}(1-\delta_{p3})
\eta_{rot}\nu_{rot}M\left[(v_{rot})_p\right]^2~~;
\end{lefteqnarray}
and the related module and trace, respectively,
read:
\begin{lefteqnarray}
\label{eq:J}
&& J=\eta_{anm}\nu_{anm}M(1+\epsilon_{21}^2)
a_1(v_{rot})_1~~; \\
\label{eq:Er}
&& E_{rot}=\eta_{rot}\nu_{rot}M(1+\epsilon_
{21}^2)\left[(v_{rot})_1\right]^2~~;
\end{lefteqnarray}
where $\eta_{anm}\nu_{anm}$, $\eta_{rot}$,
are shape factors, $\nu_{amn}$, $\nu_{rot}$,
are profile factors, $(v_{rot})_p$ is the
rotational velocity at the end of the
semiaxis, $a_p$, $p=1,2$, and the rotation
axis has been chosen to be $x_3$.   For
further details, see CM03, CM05.

The combination of Eqs.\,(\ref{eq:J}) and 
(\ref{eq:Er}) yields:
\begin{leftsubeqnarray}
\slabel{eq:ErJpqa}
&& (E_{rot})_{pq}=\frac {J^2}{Ma_1^2}\nu_{ram}(B_
{ram})_{pq}=\frac {J^2}{Ma_1^2}{\cal R}_{pq}~~; \\
\slabel{eq:ErJpqb}
&& \nu_{ram}=\frac{\nu_{rot}}{\nu_{anm}^2}~~; \\
\slabel{eq:ErJpqc}
&& (B_{ram})_{pq}=\delta_{pq}(1-\delta_{p3})\frac
{\eta_{rot}}{\eta_{anm}^2}\frac{\epsilon_{p1}^2}
{(1+\epsilon_{21}^2)^2}~~; \\
\slabel{eq:ErJpqd}
&& {\cal R}_{pq}=\nu_{ram}(B_{ram})_{pq}~~;
\label{seq:ErJpq}
\end{leftsubeqnarray}
which makes an alternative expression of the
rotation-energy tensor, and:
\begin{leftsubeqnarray}
\slabel{eq:ErJa}
&& E_{rot}=\frac {J^2}{Ma_1^2}\nu_{ram}B_{ram}=
\frac {J^2}{Ma_1^2}{\cal R}~~; \\
\slabel{eq:ErJb}
&& B_{ram}=\frac{\eta_{rot}}{\eta_{anm}^2}\frac1
{1+\epsilon_{21}^2}~~; \\
\slabel{eq:ErJc}
&& {\cal R}=\nu_{ram}B_{ram}~~;
\label{seq:ErJ}
\end{leftsubeqnarray}
which makes an alternative expression of the
rotation energy.

The limiting situations outlined above, read:
\begin{equation}
\label{eq:hrr}
\nu_{anm}=\nu_{rot}=\nu_{inr}=\frac15~~;\qquad
\nu_{ram}=\nu_{inr}^{-1}=5~~;
\end{equation}
for homogeneous configurations in rigid 
rotation, according to case (a);
\begin{equation}
\label{eq:hep}
\nu_{anm}=\frac14~~;\qquad\nu_{rot}=\frac13~~;
\qquad\nu_{ram}=\frac{16}3~~;
\end{equation}
for homogeneous configurations with constant
velocity on the equatorial plane, according
to cases (b) and (c);
\begin{equation}
\label{eq:rri}
\eta_{anm}=1~~;\qquad\eta_{rot}=\frac12~~;
\end{equation}
for rigidly rotating isopycnic surfaces,
cases (a) and (c);
\begin{equation}
\label{eq:cve}
\eta_{anm}=\frac{3\pi}8~~;\qquad\eta_{rot}=\frac34~~;
\end{equation}
for constant velocity everywhere, case (b).
Further details can be found in CM03, CM05.

\subsection{Virial equilibrium configurations}
\label{vire}

The virial theorem holds for mass distributions
which always remain confined within a finite
region of space, and the virial equations are 
satisfied, with regard to values of 
parameters averaged over a sufficiently 
long time, $t\gg T$, where $T$ is a 
characteristic period of the system (e.g.,
Landau \& Lifchitz 1966, Chap.\,II, \S\,10).
Let us define virial equilibrium as
characterized by the validity of the virial
theorem, and relaxed and unrelaxed
configurations as systems where virial 
equilibrium does and does not coincide,
respectively, with dynamical (or 
hydrostatic) equilibrium (CM05).  

Let us define systematic and peculiar
motions as related to velocity fields
where the mean value is different from
or equal to zero, respectively.   For instance,
star motions may safely be conceived as
systematic in galactic disks, and peculiar
in galactic spheroids.   In the formulation
of the virial equations, systematic motions 
related to axial rotation shall explicitly be
treated, and the remaining contributions
(e.g., vorticity, streaming, peculiar motions)
shall globally be considered together, and 
denoted as ``residual motions''.   If peculiar
motions may safely be thought of as dominant,
apart from systematic rotation, then in the
following ``residual'' has to be intended as
``peculiar''.   

With regard to unrelaxed configurations,
the generalized tensor virial equations
read (CM05):
\begin{eqnarray}
\label{eq:virte}
&& (E_{sel})_{pq}+2(E_{rot})_{pq}+2
\zeta_{pq}E_{res}=0~~; \\
\label{eq:zepq}
&& \zeta_{pq}=\frac{(\tilde{E}_{res})_{pq}}
{E_{res}}~~;\qquad p=1,2,3~~;\qquad
q=1,2,3~~; \\
\label{eq:zeta}
&&\sum_{p=1}^3\zeta_{pp}=\frac{\tilde{E}_
{res}}{E_{res}}=\zeta~~;\qquad
0\le\zeta_{pp}\le\zeta~~; \\
\label{eq:zpq0}
&& \zeta_{pq}=0~~;\qquad p\ne q~~;
\end{eqnarray}
where $\zeta_{pq}$ is the generalized
anisotropy tensor, $E_{res}$ is the
residual energy, and $(\tilde{E}_{res})
_{pq}=\zeta_{pq}E_{res}$ is the effective
residual-energy tensor i.e. the right
amount needed for the configuration of
interest to be relaxed.   The diagonal
components of the generalized anisotropy
tensor, $\zeta_{pp}$, may be conceived as
generalized anisotropy parameters.   The
related trace, $\zeta$, may be conceived
as a virial index, where $\zeta=1$
corresponds to null virial excess,
$2\Delta E_{res}=2(\tilde{E}_{res}-E_{res})$,
which does not necessarily imply a relaxed
configuration
\footnote{For instance, a homogeneous sphere    
undergoing coherent oscillations exhibits
$\zeta>1$ at expansion turnover and $\zeta<1$
at contraction turnover.  Then it necessarily
exists a configuration where $\zeta=1$ which,
on the other hand, is unrelaxed.}, %
$\zeta>1$ to positive virial excess, and
$\zeta<1$ to negative virial excess.

Let us define the effective anisotropy tensor, as:
\begin{lefteqnarray}
\label{eq:tzpq}
&& \tilde{\zeta}_{pq}=\frac{(\tilde{E}_{res})_{pq}}
{\tilde{E}_{res}}~~;\qquad p=1,2,3~~;\qquad
q=1,2,3~~; \\
\label{eq:tzet}
&&\sum_{p=1}^3\tilde{\zeta}_{pp}=1~~;\qquad
0\le\tilde{\zeta}_{pp}\le1~~; \\
\label{eq:tzp0}
&& \tilde{\zeta}_{pq}=0~~;\qquad p\ne q~~;
\end{lefteqnarray}
where the diagonal components, $\tilde{\zeta}_
{pp}$, may be conceived as effective anisotropy
parameters.
The combination of Eqs.\,(\ref{eq:zepq}), (\ref
{eq:zeta}), and (\ref{eq:tzpq}) yields:
\begin{equation}
\label{eq:zaze}
\tilde{\zeta}_{pp}=\frac{\zeta_{pp}}\zeta~~;\qquad
p=1,2,3~~;
\end{equation}
and the specification of the effective and
generalized anisotropy parameters,
$\tilde{\zeta}_{pp}$ and $\zeta_{pp}$, and the
residual kinetic energy, $E_{res}$, implies
the specification of the effective, residual
kinetic-energy tensor, $(\tilde{E}_{res})_
{pq}$.

The substitution of Eqs.\,(\ref{eq:Espq})-(\ref
{eq:Spq}) and (\ref{seq:ErJpq})-(\ref{seq:ErJ})
into the virial equations (\ref{eq:virte}), and
the particularization to the rotation axis, $p=3$,
allows the following expression of the residual 
energy (CM05):
\begin{equation}
\label{eq:Epec}
E_{res}=\frac12\frac{GM^2}{a_1}\frac{{\cal S}_{33}}
{\zeta_{33}}~~;
\end{equation}
accordingly, the remaining virial equations
read (CM05):
\begin{lefteqnarray}
\label{eq:zShR}
&& (\zeta_{33}{\cal S}_{qq}-\zeta_{qq}{\cal
S}_{33})-2h\zeta_{33}{\cal R}_{qq}=0~~;
\qquad q=1,2~~; \\
\label{eq:h}
&& h=\frac{J^2}{GM^3a_1}~~;
\end{lefteqnarray}
where $h$ may be conceived as a rotation
parameter.   It is apparent that
Eq.\,(\ref{eq:zShR}) admits real solutions
provided the following inequalities hold:
\begin{equation}
\label{eq:cone}
\frac{\zeta_{qq}}{\zeta_{33}}\le\frac{{\cal S}_
{qq}}{{\cal S}_{33}}~~;\qquad q=1,2~~;
\end{equation}
which is the natural extension of its counterpart
related to axisymmetric, relaxed configurations 
(Wiegandt1982a,b).

\subsection{Rotation and anisotropy parameters}
\label{rapa}

To get a more evident connection with the
centrifugal potential on the boundary, let
us define the rotation parameter (CM05):
\begin{equation}
\label{eq:vg}
\upsilon=\frac{\Omega^2}{2\pi G\bar{\rho}}~~;
\end{equation}
where $\Omega=\Omega(a_1,0)$ is the angular
velocity at the end of the major equatorial semiaxis,
denoted as $a_1$, and $\bar{\rho}$ is the mean density
of the ellipsoid:
\begin{equation}
\label{eq:rome}
\bar{\rho}=\frac3{4\pi}\frac M{a_1a_2a_3}~~;
\end{equation}
the above definition of the rotation parameter,
$\upsilon$, makes a generalization of some
special cases mentioned in the literature
(e.g., Jeans 1929, Chap.\,IX, \S 232; 
Chandrasekhar \& Leboviz 1962).

The combination of Eqs.\,(\ref{eq:J}),
(\ref{eq:h}), (\ref{eq:vg}), and
(\ref{eq:rome}) yields:
\begin{equation}
\label{eq:hups}
h=\frac32\eta_{anm}^2\nu_{anm}^2\frac{(1+\epsilon_{21}^2
)^2}{\epsilon_{21}\epsilon_{31}}\upsilon~~;
\end{equation}
which links the rotation parameters, $h$ and
$\upsilon$.   An explicit expression of the
rotation parameter, $h$, may directly be
obtained from Eq.\,(\ref{eq:zShR}), as:
\begin{equation}
\label{eq:virh}
h=\frac12\frac{\zeta_{33}{\cal S}_{qq}-\zeta
_{qq}{\cal S}_{33}}{\zeta_{33}{\cal R}_{qq}}
~~;\quad q=1,2~~;
\end{equation}
and the substitution of Eqs.\,(\ref{eq:virh})
into (\ref{eq:hups}), using (\ref{seq:ErJpq})
and (\ref{seq:ErJ}), allows the explicit
expression of the rotation parameter,
$\upsilon$, as:
\begin{equation}
\label{eq:upsh}
\upsilon=\frac13
\frac{\epsilon_{21}\epsilon_{31}}{(1+\epsilon_
{21}^2)^2}\frac1{\eta_{amn}^2\nu_{amn}^2}\frac
{\zeta_{33}{\cal S}_{qq}-\zeta_{qq}{\cal S}_{33}}
{\zeta_{33}{\cal R}_{qq}}~~;\quad q=1,2~~;
\end{equation}
which, in the special case of rigidly
rotating, homogeneous configurations
with isotropic peculiar velocity 
distribution, reduces to a known
relation for Jacobi ellipsoids and,
with the additional demand of axial
symmetry, to a known relation for
MacLaurin spheroids (e.g., Jeans
1929, Chap.\,VIII, \S\S 189-193;
Chandrasekhar 1969, Chap.\,5, \S
32, Chap.\,6, \S 39; Caimmi 1996a).

An explicit expression of anisotropy
parameter ratio, can be obtained via
Eq.\,(\ref{eq:virh}).   The result is
(CM05):
\begin{lefteqnarray}
\label{eq:zq3}
&& \frac{\zeta_{qq}}{\zeta_{33}}=\frac
{{{\cal S}}_{qq}}{{{\cal S}}_{33}}
\left[1-2h\frac{{{\cal R}}_{qq}}{{{\cal
S}}_{qq}}\right]~~;\quad q=1,2~~; \\
\label{eq:z12}
&& \frac{\zeta_{11}}{\zeta_{22}}=\frac
{{{\cal S}}_{11}-2h{{\cal R}}_{11}}
{{{\cal S}}_{22}-2h{{\cal R}}_{22}}~~;
\end{lefteqnarray}
and the combination of Eqs.\,(\ref{eq:zeta}) 
and (\ref{eq:zq3}) yields:
\begin{equation}
\label{eq:z3}
\frac{\zeta_{33}}{\zeta}=\frac{{{\cal S}}_{33}}
{{{\cal S}}-2h{{\cal R}}}~~;
\end{equation}
which provides an alternative expression to
Eqs.\,(\ref{eq:virh}) and (\ref{eq:upsh}), as:
\begin{lefteqnarray}
\label{eq:hz}
&& h=\frac12\frac{\zeta_{33}{{\cal S}}-\zeta
{{\cal S}}_{33}}{\zeta_{33}{{\cal R}}}~~; \\
\label{eq:upz}
&& \upsilon=\frac13\frac{\epsilon_{21}\epsilon_
{31}}{(1+\epsilon_{21}^2)^2}\frac1{\eta_{amn}^2\nu_
{amn}^2}\frac{\zeta_{33}{{\cal S}}-\zeta
{{\cal S}}_{33}}{\zeta_{33}{{\cal R}}}~~;
\end{lefteqnarray}
that are equivalent to Eq.\,(\ref{eq:zShR}), and
then admit real solutions provided inequality
(\ref{eq:cone}) is satisfied.

Finally, Eqs.\,(\ref{eq:zShR}) may be combined as:
\begin{equation}
\label{eq:R12}
\frac{{{\cal R}}_{11}}{{{\cal R}}_{22}}=\frac
{\zeta_{33}{{\cal S}}_{11}-\zeta_{11}
{{\cal S}}_{33}}{\zeta_{33}{{\cal S}}_{22}-
\zeta_{22}{{\cal S}}_{33}}~~;
\end{equation}
where it can be seen that Eqs.\,(\ref
{eq:z12}) and (\ref{eq:R12}) are changed
one into the other, by replacing the
terms, ${{\cal S}}_{33}\zeta_{qq}/\zeta_
{33}$, with the terms, $2h{{\cal R}}_{qq}$,
and vice versa.   The above results may be
reduced to a single statement.
\begin{trivlist}
\item[\hspace\labelsep{\bf Theorem 1.}] \sl
Given a homeoidally striated Jacobi ellipsoid,
rotating around an axis, $x_3$, the relation:
\begin{lefteqnarray*}
&& \frac{X_{11}}{X_{22}}=\frac{{{\cal S}}_{11}
-Y_{11}}{{{\cal S}}_{22}-Y_{22}}~~; \\
&& X_{qq}=\frac{\zeta_{qq}}{\zeta_{33}}{{\cal
S}}_{33}~~,\quad2h{{\cal R}}_{qq}~~;\qquad
q=1,2~~; \\
&& Y_{qq}=2h{{\cal R}}_{qq}~~,\quad\frac
{\zeta_{qq}}{\zeta_{33}}{{\cal S}}_{33}~~;
\qquad q=1,2~~;
\end{lefteqnarray*}
is symmetric with respect to $X_{qq}$ and
$Y_{qq}$, being the former tensor related
to anisotropic residual velocity distribution,
and the latter one to systematic rotation
around the axis, $x_3$, or vice versa.
\end{trivlist}

In the special case of axisymmetric configurations,
the shape factors related to equatorial axes
coincide, $A_2=A_1$, as $\epsilon_{21}=1$ (e.g.,
CM05).   Then ${{\cal S}}_{11}={{\cal S}}_{22}$
owing to Eqs.\,(\ref{eq:Bspq}), (\ref{eq:Spq}),
and ${{\cal R}}_{11}={{\cal R}}_{22}$ owing to
Eqs.\,(\ref{seq:ErJ}), which
necessarily imply $\zeta_{11}=\zeta_{22}$,
owing to Eq.\,(\ref{eq:R12}).

In the general case of triaxial configurations,
the contrary holds, $A_2\ne A_1$, as $\epsilon_
{21}\ne1$, then ${{\cal S}}_{11}\ne{{\cal S}}_
{22}$ and ${{\cal R}}_{11}\ne{{\cal R}}_{22}$,
then the equality,
$\zeta_{11}=\zeta_{22}$, via Eq.\,(\ref
{eq:z12}), implies the validity of the relation:
\begin{equation}
\label{eq:hiseq}
h=\frac12\frac{{\cal S}_{11}-{\cal S}_{22}}
{{\cal R}_{11}-{\cal R}_{22}}~~;
\end{equation}
if otherwise, the residual velocity distribution
along the equatorial plane
\footnote{Throughout this paper, ``along the
equatorial plane'' has to be intended as
``along any direction parallel to the equatorial
plane''.}
is anisotropic i.e.
$\zeta_{11}\ne\zeta_{22}$.   The related degeneracy
can be removed using an additional condition,
as it will be shown in the next section.

The above results may be reduced to a single
statement.
\begin{trivlist}
\item[\hspace\labelsep{\bf Theorem 2.}] \sl
Isotropic residual velocity distribution
along the equatorial plane, $\zeta_{11}=
\zeta_{22}$, makes a necessary condition
for homeoidally striated Jacobi ellipsoids
to be symmetric with respect to the
rotation axis, $x_3$, i.e. reduce to
homeoidally striated MacLaurin spheroids.
\end{trivlist}

\section{A unified theory of systematic
and random motions}
\label{unte}
\subsection{Imaginary rotation}
\label{imro}

A unified theory of systematic and random
motions is allowed, taking into consideration
imaginary rotation.   It has been shown above
that Eq.\,(\ref{eq:zShR}), or equivalently
one among (\ref{eq:virh}), (\ref{eq:upsh}),
(\ref{eq:hz}), (\ref{eq:upz}), admits real
solutions provided inequality (\ref{eq:cone})
is satisfied.   If otherwise, the rotation
parameter - let it be $h$ or $\upsilon$ -
has necessarily to be negative, which implies,
via Eqs.\,(\ref{eq:h}) or (\ref{eq:vg}), an
{\it imaginary} angular velocity, $\img\Omega$,
where $\img$ is the imaginary unit.
Accordingly, the centrifugal potential
takes the general expression:
\begin{equation}
\label{eq:Tpm}
{\cal T}^\mp(x_1,x_2,x_3)=\mp\frac12[\Omega
(x_1,x_2,x_3)]^2w^2~~;\qquad w^2=x_1^2+x_2^2~~;
\end{equation}
where the minus and the plus correspond to
imaginary and real rotation, respectively.
The centrifugal force related to real
rotation, $\partial{\cal T}^+/\partial w$,
has opposite sign with respect to the
gravitational force, $\partial{\cal V}/
\partial w$.   On the
other hand, the centrifugal force related
to imaginary rotation, $\partial{\cal T}^-/
\partial w$, has equal sign with respect to
the gravitational force, $\partial{\cal V}/
\partial w$.   Then the net effect
of real rotation is flattening, while the net
effect of imaginary rotation is {\it elongation},
with respect to the rotation axis (Caimmi 1996b).

To get further insight, let us particularize
Eq.\,(\ref{eq:zShR}) to the special case of
null rotation $(h=0)$.   The result is:
\begin{equation}
\label{eq:zq30}
\frac{\zeta_{qq}}{\zeta_{33}}=\frac{{{\cal S}}_{qq}}
{{{\cal S}}_{33}}~~;\qquad h=0~~;\qquad q=1,2~~;
\end{equation}
where the right-hand side, via Eqs.\,(\ref{eq:Bspq})
and (\ref{eq:Spq}), depends on the axis ratios only.
It can be seen (Appendix A) that ${{\cal S}}_{qq}/
{{\cal S}}_{33}\ge1$ for oblate-like configurations
$(a_1\ge a_2\ge a_3)$, and ${{\cal S}}_{qq}/
{{\cal S}}_{33}\le1$ for prolate-like configurations
$(a_2\le a_1\le a_3)$, which implies $\zeta_{qq}/
\zeta_{33}\ge1$ for oblate-like configurations,
and $\zeta_{qq}/\zeta_{33}\le1$ for prolate-like
configurations.   Accordingly, the net effect of
positive or negative
residual motion excess along the equatorial
plane is flattening or elongation, respectively.
In what follows, it shall be intended that
residual motion excess is related to the
equatorial plane.

The above considerations show a strict connection
between real rotation and positive residual motion
excess, on one hand, and between imaginary rotation
and negative residual motion excess, on the other
hand.   At this point, it is natural to ask if
(positive or negative) residual motion excess
can be interpreted as (real or imaginary) rotation,
and vice versa.

\subsection{Residual motion excess and rotation}
\label{preq}

With regard to the rotation parameter, $\upsilon$,
it is convenient to use the more compact notation
\footnote{The factor, 3, does not appear in     
CM05: $\upsilon_N=3(\upsilon_N)_{CM05}$.}:%
\begin{equation}
\label{eq:upzN}
\upsilon_N=\frac{3\eta_{rot}\nu_{rot}}
{\nu_{sel}}\upsilon~~;
\end{equation}
and the substitution of Eqs.\,(\ref{eq:Bspq}),
(\ref{eq:Spq}), (\ref{seq:ErJpq}), (\ref
{seq:ErJ}), (\ref{eq:zaze}), into (\ref{eq:upsh})
and (\ref{eq:upz}), yields the equivalent
expressions:
\begin{lefteqnarray}
\label{eq:uNq3}
&& \upsilon_N=A_q-\frac{\zeta_{qq}}{\zeta_{33}}
\epsilon_{3q}^2A_3~~;\qquad q=1,2~~; \\
\label{eq:uNq}
&& \upsilon_N=\frac1{1+\epsilon_{21}^2}\left[
A_1+\epsilon_{21}^2A_2+\frac{\zeta_{33}-\zeta}
{\zeta_{33}}\epsilon_{31}^2A_3\right]~~;
\end{lefteqnarray}
which shows that, for homeoidally striated
Jacobi ellipsoids, the normalized rotation
parameter, $\upsilon_N$, depends on the axis
ratios, $\epsilon_{21}$, $\epsilon_{31}$,
and a {\it single} anisotropy parameter,
$\tilde{\zeta}_{33}=\zeta_{33}/\zeta$.
In the special case of homogeneous, rigidly
rotating configurations, owing to Eqs.\,(\ref
{eq:omogb}), (\ref{eq:hrr}), and (\ref{eq:rri}),
Eq.\,(\ref{eq:upzN}) reduces to: $\upsilon_N=
\upsilon$.

In the limit of isotropic residual velocity
distribution, $\zeta_{11}=\zeta_{22}=\zeta_
{33}=\zeta/3$, Eqs.\,(\ref{eq:uNq3}) and
(\ref{eq:uNq}) take the simpler form:
\begin{lefteqnarray}
\label{eq:uq3i}
&& (\upsilon_N)_{iso}=A_q-
\epsilon_{3q}^2A_3~~;\qquad q=1,2~~; \\
\label{eq:uqi}
&& (\upsilon_N)_{iso}=\frac1{1+\epsilon_{21}^2}\left[
A_1+\epsilon_{21}^2A_2-2\epsilon_{31}^2A_3\right]~~;
\end{lefteqnarray}
where the index, $iso$, means isotropic
residual velocity distribution.

Accordingly, Eqs.\,(\ref{eq:uNq3}) and (\ref{eq:uNq})
may be expressed as:
\begin{lefteqnarray}
\label{eq:uNia}
&& \upsilon_N=(\upsilon_N)_{iso}-(\upsilon_N)_
{ani}~~; \\
\label{eq:uq3a}
&& (\upsilon_N)_{ani}=\left(\frac{\zeta_{qq}}
{\zeta_{33}}-1\right)\epsilon_{3q}^2A_3~~; \\
\label{eq:u3a}
&& (\upsilon_N)_{ani}=\left(\frac\zeta{\zeta_{33}}-3
\right)\frac{\epsilon_{31}^2A_3}{1+\epsilon_{21}^
2}~~;
\end{lefteqnarray}
where $(\upsilon_N)_{ani}\ge0$ for oblate-like
configurations, $\zeta_{qq}/\zeta_{33}\ge1$;
$(\upsilon_N)_{ani}\le0$ for prolate-like
configurations, $\zeta_{qq}/\zeta_{33}\le1$;
and the index, $ani$, means contribution from
residual motion excess.   Accordingly, positive
residual motion excess along the equatorial plane
is related to real rotation.   On the contrary,
negative residual motion excess along the rotation
axis is related to imaginary rotation.   

Let us rewrite Eq.\,(\ref{eq:uNia}) as:
\begin{equation}
\label{eq:uNo}
(\upsilon_N)_{iso}=\upsilon_N+(\upsilon_N)_
{ani}~~;
\end{equation}
which, owing to Eqs.\,(\ref{eq:vg}) and
(\ref{eq:upzN}), is equivalent to:
\begin{equation}
\label{eq:O2i}
\Omega^2_{iso}=\Omega^2\mp\Omega^2_{ani}~~;
\end{equation}
where the plus corresponds to real rotation,
$(\upsilon_N)_{ani}\ge0$, and the minus to
imaginary rotation, $(\upsilon_N)_{ani}\le0$.
Then the effect of residual motion excess on
the shape
of the system, is virtually indistinguishible
from the effect of additional systematic
rotation, with similar (rotation) velocity
distribution as the pre-existing one.  The
related, explicit expression, according to
Eqs.\,(\ref{eq:rvang}), (\ref{eq:vg}), and
(\ref{eq:u3a}), is:
\begin{lefteqnarray}
\label{eq:OaO}
&& \frac{\Omega_{ani}(r,\theta)}{\Omega_{ani}
(R,\theta)}=\frac{\Omega_{ani}(a_1^\prime,0)}
{\Omega_{ani}(a_1^\prime,0)}=\frac{\Omega(a_1^
\prime,0)}{\Omega(a_1^\prime,0)}=\frac{\Omega
(r,\theta)}{\Omega(R,\theta)}~~; \\
\label{eq:O2a}
&& \Omega^2_{ani}=\Omega^2_{ani}(a_1)=\mp
(\upsilon_N)_{ani}2\pi G\overline{\rho}~~;
\end{lefteqnarray}
where the double sign ensures a non negative
value on the right hand-side member of
Eq.\,(\ref{eq:O2a}).  The above
results may be reduced to a single statement.
\begin{trivlist}
\item[\hspace\labelsep{\bf Theorem 3.}] \sl
Given a homeoidally striated Jacobi ellipsoid,
the effect of (positive or negative) residual
motion excess is
equivalent to an additional (real or imaginary)
rotation, with equal spatial distribution as
the pre-existing rotation, but value at the
end of major equatorial axis conform to the
relation:
\begin{equation}
\label{eq:Ot3}
\mp\Omega^2_{ani}=\mp\Omega^2_{ani}(a_1)=2\pi G
\overline{\rho}\left(\frac\zeta{\zeta_{33}}-3
\right)\frac{\epsilon_{31}^2A_3}{1+\epsilon_
{21}^2}~~;
\end{equation}
with regard to an adjoint configuration
where the residual velocity distribution 
is isotropic.
\end{trivlist}
Accordingly, a homeoidally striated Jacobi
ellipsoid with assigned systematic rotation
and residual velocity distribution, with
regard to the shape, is virtually
indistinguishable from an adjoint
configuration of equal mass and axes,
systematic rotation velocity distribution
deduced from Eqs.\,(\ref{eq:rvang}),
(\ref{eq:OaO}) and (\ref{eq:O2a}), and
isotropic residual velocity distribution.

Owing to Eqs.\,(\ref{eq:uNia}), (\ref{eq:uq3a}),
and (\ref{eq:u3a}), Eqs.\,(\ref{eq:uq3i}) and
(\ref{eq:uqi}) are equivalent to (\ref{eq:uNq3})
and (\ref{eq:uNq}), respectively.   On the other
hand, Eqs.\,(\ref{eq:uq3i}) and (\ref{eq:uqi})
are valid for classical Jacobi ellipsoids with
rotation parameter, $\upsilon=(\upsilon_N)_{iso}$.
The above results may be reduced to a single statement.
\begin{trivlist}
\item[\hspace\labelsep{\bf Theorem 4.}] \sl
Given a homeoidally striated Jacobi ellipsoid,
the adjoint configuration, characterized by
isotropic residual velocity distribution and
normalized rotation parameter, $(\upsilon_N)_
{iso}$, exhibits same shape as a classical
Jacobi ellipsoid of equal mass and axes, and
rotation parameter, $\upsilon=(\upsilon_N)_
{iso}$.
\end{trivlist}
Accordingly, a homeoidally striated Jacobi
ellipsoid (in general, an unrelaxed configuration)
with assigned systematic rotation and residual
velocity distribution, with regard to the shape,
is virtually indistinguishable from an adjoint,
classical Jacobi ellipsoid (a relaxed
configuration) of equal mass and axes, and
rotation parameter equal to the normalized
rotation parameter of the original configuration.

\subsection{Axis ratios and anisotropy parameters}
\label{neco}

The combination of alternative expressions of
the rotation parameter, $(\upsilon_N)_{iso}$,
expressed by Eqs.\,(\ref{eq:uq3i}), yields:
\begin{equation}
\label{eq:upep}
\epsilon_{21}^2(A_2-A_1)=(1-\epsilon_{21}^2)
\epsilon_{31}^2A_3~~;
\end{equation}
which, for axisymmetric configurations,
reduces to an indeterminate form, $0=0$.

The combination of alternative expressions of
the rotation parameter, $(\upsilon_N)_{ani}$,
expressed by Eqs.\,(\ref{eq:uq3a}), yields:
\begin{equation}
\label{eq:ziep}
\zeta_{33}-\zeta_{22}=\epsilon_{21}^2(\zeta_
{33}-\zeta_{11})~~;
\end{equation}
which, for isotropic residual velocity
distributions, reduces to an indeterminate
form, $0=0$.   In addition, axisymmetric
configurations $(\epsilon_{21}=1)$
necessarily imply isotropic residual
velocity distributions along the equatorial
plane, $\zeta_{11}=\zeta_{22}$.

The combination of Eqs.\,(\ref{eq:zeta})
and (\ref{eq:ziep}) yields:
\begin{leftsubeqnarray}
\slabel{eq:z12a}
&& \zeta_{11}=\frac{\zeta-(2-\epsilon_{21}^2)\zeta_
{33}}{1+\epsilon_{21}^2}~~; \\
\slabel{eq:z12b}
&& \zeta_{22}=\frac{\epsilon_{21}^2\zeta+(1-2
\epsilon_{21}^2)\zeta_{33}}{1+\epsilon_{21}^2}~~;
\label{seq:z12}
\end{leftsubeqnarray}
which, for axisymmetric configurations
$(\epsilon_{21}=1)$ reduces to $\zeta_
{11}=\zeta_{22}=(\zeta-\zeta_{33})/2$,
and the special case, $\zeta_{33}=\zeta/3$,
reads $\zeta_{11}=\zeta_{22}=\zeta/3$.

The conditions, $\zeta_{11}\ge0$, related
to Eqs.\,(\ref{eq:zepq}) and (\ref{eq:zeta}),
and $0\le\epsilon_{21}\le1$, necessarily
imply $0\le\zeta_{33}/\zeta\le1/2$.   If
otherwise, sequences of virial equilibrium
configurations cannot attain the oblong
shape, $\epsilon_{21}=\epsilon_{31}=0$,
but must stop earlier, where the equivalent
relations:
\begin{equation}
\label{eq:epzi}
\epsilon_{21}^2=\frac{2\zeta_{33}-\zeta}
{\zeta_{33}}~~;\qquad\zeta_{11}=0~~;
\end{equation}
are satisfied.   The above results may
be reduced to the following statements.
\begin{trivlist}
\item[\hspace\labelsep{\bf Theorem 5.}] \sl
Given a homeoidally striated Jacobi ellipsoid,
the anisotropy parameters along the equatorial
plane depend on the equatorial axis ratio for
triaxial configurations.   The related expressions
coincide, $\zeta_{11}=\zeta_{22}$, in the limit
of axisymmetric configurations, $\epsilon_{21}=1$.
\end{trivlist}
\begin{trivlist}
\item[\hspace\labelsep{\bf Theorem 6.}] \sl
A necessary and sufficient condition for a
homeoidally striated Jacobi ellipsoid to
have isotropic residual velocity distribution,
is that the anisotropy parameter along the
rotation axis, equals one third.
\end{trivlist}
\begin{trivlist}
\item[\hspace\labelsep{\bf Theorem 7.}] \sl
Given a sequence of homeoidally striated Jacobi
ellipsoids, the ending point occurs when the
equatorial axis ratio is null, $\epsilon_{31}=
0$, and/or the generalized anisotropy parameter
related to the major axis is null, $\zeta_{11}
=0$, which is equivalent to $\epsilon_{21}=(2-
\zeta/\zeta_{33})^{1/2}$.   The special case
of dynamical (or hydrostatic) equilibrium,
$\zeta=1$, implies centrifugal support along
the major axis provided $\zeta_{11}=0$.
\end{trivlist}

Accordingly, with regard to homeoidally
striated Jacobi ellipsoids, the anisotropy
parameters along the equatorial plane,
$\zeta_{11}$ and $\zeta_{22}$, cannot be
arbitrarily assigned, but depend on the
equatorial axis ratio, $\epsilon_{21}$,
conform to Eqs.\,(\ref{seq:z12}).   On
the other hand, the further knowledge of the
meridional axis ratio, $\epsilon_{31}$,
and the rotation parameter, $\upsilon_N$,
allows the determination of the rotation
parameter, $(\upsilon_N)_{ani}$, via
Eqs.\,(\ref{eq:uq3i}), (\ref{eq:uqi}),
(\ref{eq:uNia}), and then the ratios,
$\zeta_{qq}/\zeta_{33}$, $\zeta_{33}/
\zeta$, via Eqs.\,(\ref{eq:uq3a}),    
(\ref{eq:u3a}), respectively, or the
anisotropy parameter
along the rotation axis, $\zeta_{33}$,
provided the virial index, $\zeta$,
defined by Eq.\,(\ref{eq:zeta}), is
assigned.

In conclusion, with regard to homeoidally
striated Jacobi ellipsoids defined by
assigned axis ratios, $\epsilon_{31}$,
$\epsilon_{21}$, rotation parameter,
$\upsilon_N$, and virial index, $\zeta$,
the anisotropy parameters, $\zeta_{11}$,
$\zeta_{22}$, $\zeta_{33}$, cannot
arbitrarily be fixed, but must be
determined as shown above.

\subsection{Sequences of virial equilibrium
configurations}
\label{seve}

With regard to homeoidally striated Jacobi
ellipsoids, it has been shown above that
adjoints configurations are characterized
by (i) centrifugal potential, ${{\cal T}}_
{iso}(x_1,x_2,x_3)={{\cal T}}(x_1,x_2,x_3)
+{{\cal T}}_{ani}(x_1,x_2,x_3)$, or $\Omega
^2_{iso}(x_1,x_2,x_3)=\Omega^2(x_1,x_2,x_3)
\mp\Omega^2_{ani}(x_1,x_2,x_3)$, where the angular
velocity distribution related to residual
motion excess, is defined by Eqs.\,(\ref
{eq:OaO}) and (\ref{eq:Ot3}); and (ii)
isotropic residual velocity distribution.
Owing to Theorem 3, a sequence of homeoidally
striated Jacobi ellipsoids in virial
equilibrium coincides with the sequence
of adjoints configurations.  The last, in turn,
owing to Theorem 4, coincides with the
sequence of classical Jacobi ellipsoids
provided the related rotation parameters
coincide, $\upsilon=(\upsilon_N)_{iso}$,
conform to Eqs.\,(\ref{eq:uq3i}), (\ref
{eq:uqi}), and imaginary rotation is
also considered.   Given a homeoidally
striated Jacobi ellipsoid with fixed
axis ratios, $\epsilon_{21}$, $\epsilon_
{31}$, rotation parameter, $(\upsilon_N)_
{ani}$, and virial index, $\zeta$, the
anisotropy parameters, $\zeta_{11}$,
$\zeta_{22}$, $\zeta_{33}$, are
determined via
Eqs.\,(\ref{eq:uq3i})-(\ref{eq:u3a}) and
(\ref{seq:z12}).   Negative values of
the rotation parameter, $(\upsilon_N)_
{iso}$, extend the sequence of classical
MacLaurin spheroids to imaginary rotation i.e.
prolate configurations where the major
axis coincides with the rotation axis.

Once more owing to Theorem 4, the
bifurcation point of a sequence of
homeoidally striated Jacobi ellipsoids
in virial equilibrium coincides with
its counterpart along the sequence of
classical Jacobi ellipsoids, $[\epsilon
_{31},(\upsilon_N)_{iso}]=(0.582724,
0.187115)$ (e.g., Jeans 1929,
Chap.\,VIII, \S\,192; Chandrasekhar
1969, Chap.\,5, \S\,33).   Accordingly,
the axis ratio of the configuration
where the bifurcation occurs, is a
solution of the transcendental equation
(Caimmi 1996a):
\begin{equation}
\label{eq:bifJ}
\frac{\epsilon_{31}^2A_3}{A_1}=\frac{3-2
\epsilon_{31}^2}{5-4\epsilon_{31}^2}~~;
\qquad\epsilon_{21}=1~~;
\end{equation}
where, owing to Eqs.\,(\ref{eq:disAe}):
\begin{leftsubeqnarray}
\slabel{eq:diraa}
&& \frac{\epsilon_{31}^2A_3}{A_1}\le1~~;
\qquad\epsilon_{31}\le1~~; \\
\slabel{eq:dirab}
&& \frac{\epsilon_{31}^2A_3}{A_1}\ge1~~;
\qquad\epsilon_{31}\ge1~~;
\label{seq:dira}
\end{leftsubeqnarray}
and further analysis of Eq.\,(\ref
{eq:bifJ}) shows that two solutions
exist.   One is related to spherical
configurations $(\epsilon_{31}=
\epsilon_{21}=1)$, which belong
to the axisymmetric sequence $(a_1
=a_2\ne a_3)$; for this reason, it
makes a parasite solution to the
problem under discussion.   On the
contrary, the remaining solution
defines the axisymmetric configuration
where bifurcation takes place.

\begin{figure}
\centerline{\psfig{file=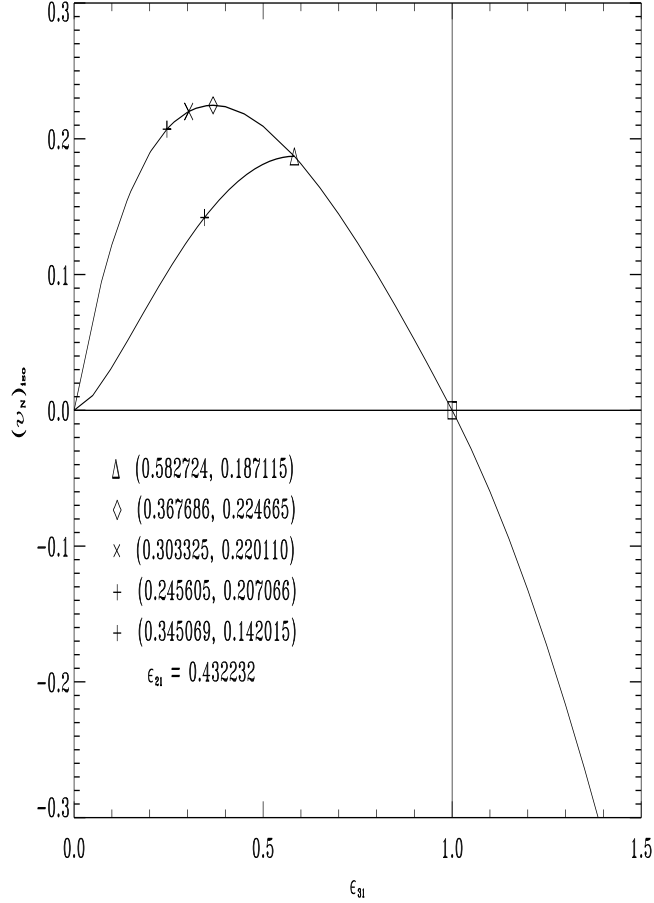,height=130mm,width=90mm}}
\caption{The sequence of classical MacLaurin
spheroids, from the starting point (square)
to the bifurcation point (triangle), and
classical Jacobi ellipsoids, from the
starting point (triangle) to the bifurcation
point (lower latin cross), extended to the
case of imaginary rotation (negative values 
of the rotation parameter, $\upsilon$, which
equals the normalized rotation parameter,
$(\upsilon_N)_{iso}$, related to homeoidally
striated Jacobi ellipsoids).   Both sequences
are continued in the region of instability.
With regard to MacLaurin spheroids, it is
also marked: the extremum point of maximum
(diamond); the point at the onset of dynamical
instability (St. Andreas cross); and the point
at the onset of instability towards pear-shaped
configurations (upper latin cross), which has
its counterpart on the Jacobi sequence.
Numerical values are taken from Chandrasekhar
(1969, Chap.\,5, \S\S\,32-33; Chap.\,6,
\S\S39-41).   The horizontal line, $(\upsilon_
N)_{iso}=0$, $\epsilon_{31}>0$, is the locus
of non rotating
configurations.   The vertical line, $\epsilon
_{31}=1$, is the locus of spherical
configurations.   The above mentioned lines
divide the non negative semiplane, $\epsilon_
{31}\ge0$, into four regions (from top left in
clockwise sense): A - oblate-like shapes
with real rotation; B - prolate-like shapes
with real rotation; C - prolate-like shapes
with imaginary rotation; D - oblate-like
shapes with imaginary rotation.   Regions
B and D are forbidden to sequences of
homeoidally striated Jacobi ellipsoids.}
\label{f:siso}
\end{figure}

\begin{figure}
\centering
\centerline{\psfig{file=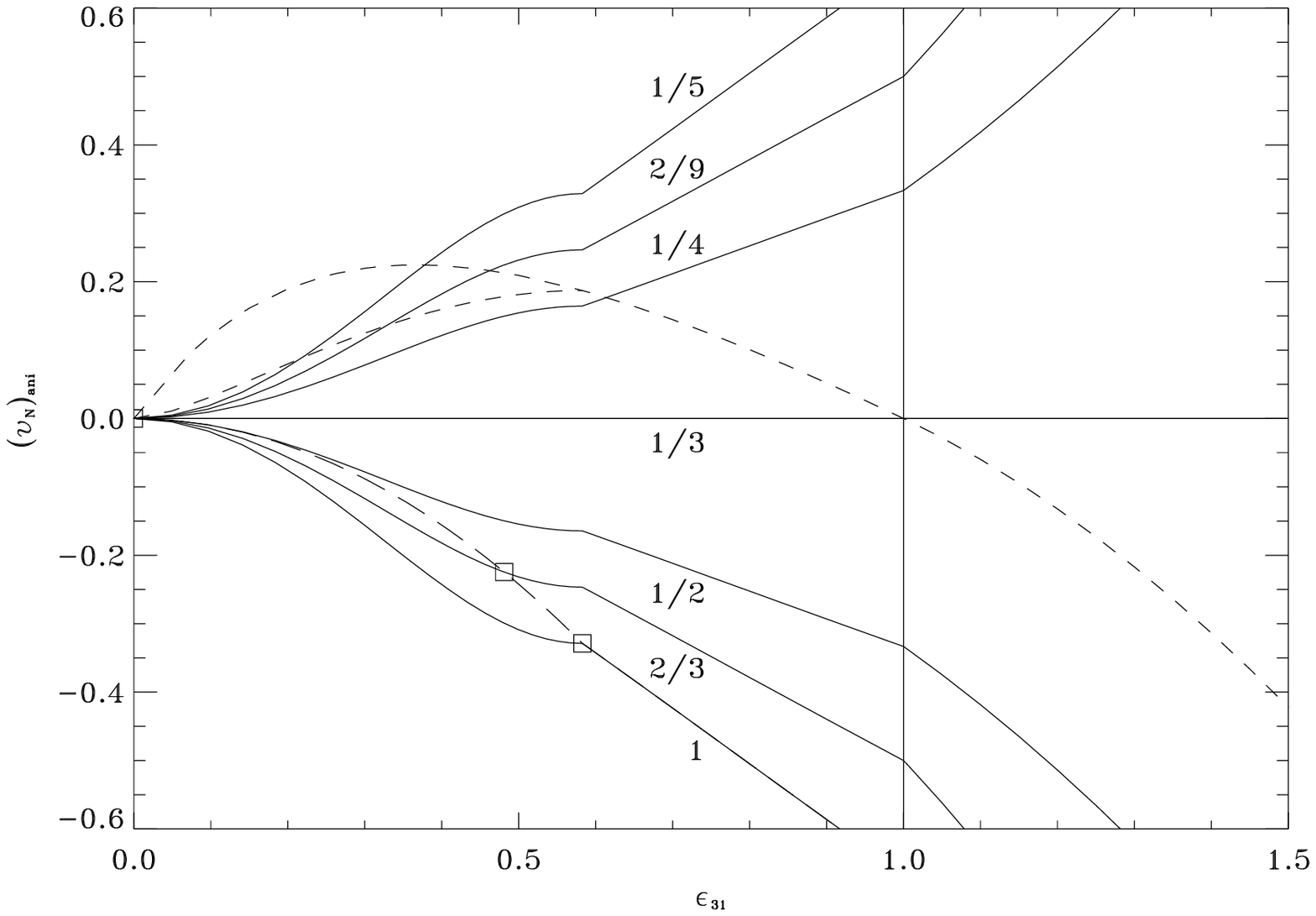,height=130mm,width=90mm}}
\caption{The normalized rotation parameter,
$(\upsilon_N)_{ani}$, related to residual
motion excess,
as a function of the axis ratio, $\epsilon_
{31}$, with regard to homeoidally striated
Jacobi ellipsoids.   Each curve is labelled
by the value of the generalized anisotropy
parameter, $\tilde{\zeta}_{33}=\zeta_{33}/
\zeta$.   The horizontal non negative
semiaxis, $\epsilon_{31}\ge0$, $(\upsilon_N)_
{ani}=0$, is the locus of configurations
with isotropic residual velocity distribution,
$\tilde{\zeta}_{33}=1/3$.   The vertical
straight line, $\epsilon_{31}=1$, is the
locus of spherical configurations.   The
generic sequence starts from a non rotating
configuration (short-dashed curves) and ends at a
configuration where either $\epsilon_{31}=0$
or $\tilde{\zeta}_{33}=0$ (long-dashed curve).
The regions above the upper short-dashed curve
and below the long-dashed curve, respectively,
are forbidden.
The non negative vertical semiaxis,
$(\upsilon_N)_{ani}\ge0$, $\epsilon_{31}=0$,
corresponds to flat $(\tilde{\zeta}_{33}=0)$
or oblong $(\tilde{\zeta}_{33}=\tilde{\zeta}
_{22}=0)$ configurations.   The curves are
symmetric with respect to the horizontal
axis, until the limiting curve, $\tilde{\zeta}_
{33}=\tilde{\zeta}=1$, is reached in region (D).
The limiting configuration
where $\tilde{\zeta}_{11}=0$, is marked by a
square: no configuration in virial
equilibrium is allowed on the left,
as it would imply $\tilde{\zeta}_{11}<0$.
No configuration is allowed on the right
of the starting point, as it would imply
$\upsilon_N<0$.   The sequence starting
from the bifurcation point (not shown)
corresponds to
$\tilde{\zeta}_{33}=0.241648$, $\tilde{\zeta}_
{11}=\tilde{\zeta}_{22}=0.379176$.}
\label{f:sani}
\end{figure}

Sequences of homeoidally striated
Jacobi ellipsoids can be deduced
from Figs.\,\ref{f:siso} and \ref
{f:sani}.   The sequence of classical
MacLaurin spheroids (including
Jacobi ellipsoids), extended to
the case of imaginary rotation,
is shown in Fig.\,\ref{f:siso}.
The rotation parameter, $(\upsilon_
N)_{ani}$, is plotted in Fig.\,\ref
{f:sani} as a function of the axis
ratio, $\epsilon_{31}$, for different
values of the effective anisotropy
parameter, $\tilde{\zeta}_{33}=
\zeta_{33}/\zeta$, labelled on
each curve.
With regard to a fixed anisotropy
parameter, $\tilde{\zeta}_{33}$,
the related sequence starts from
a non rotating configuration,
$\upsilon_N=0$ i.e. $(\upsilon_N)_
{ani}=(\upsilon_N)_{iso}$, and
ends at a configuration where either
$\epsilon_{31}=0$ or $\tilde{\zeta}_
{11}=0$.   The locus of the ending
points related to the latter
alternative, is represented as a
long-dashed curve.   The sequence cannot
be continued on the right, as
imaginary rotation i.e. larger
$\tilde{\zeta}_{33}$ would be
needed and a different sequence
should be considered.   The effect
of positive $(\tilde{\zeta}_{33}<
1/3)$ or negative $(\tilde{\zeta}_
{33}>1/3)$ residual motion excess
is equivalent to an additional
real or imaginary rotation,
respectively.
The horizontal non negative
semiaxis, $\epsilon_{31}\ge0$, $(\upsilon_N)_
{ani}=0$, is the locus of configurations
with isotropic residual velocity distribution,
$\tilde{\zeta}_{33}=1/3$.   The vertical
straight line, $\epsilon_{31}=1$, is the
locus of spherical configurations.

With regard to real rotation, the ending
point of a sequence occurs at either null
meridional axis ratio, $\epsilon_{31}=0$,
and/or null effective anisotropy parameter
related to major equatorial axis direction,
$\tilde{\zeta}_{11}=0$.   The occurrence
of the latter situation is marked by a
square on the corresponding sequence
plotted in Fig.\,\ref{f:sani}, and the
related locus is represented as a long-dashed
curve.
Configurations on the left are forbidden,
as they would imply negative residual
energy tensor component, $(E_{res})_{11}
<0$, to satisfy the virial equation (\ref
{eq:virte}), which would imply imaginary
rotation around major equatorial axis i.e.
systematic motions other than rotation
around the minor axis.

With regard to imaginary rotation, no
ending point occurs and the system is
allowed to attain negative infinite
rotation parameter, $(\upsilon_N)_{iso}
\to-\infty$, and infinite meridional
axis ratio, $\epsilon_{31}\to+\infty$.
The related configuration is either a
spindle $(a_1=a_2=0)$ or an infinitely
high cylinder $(a_1=a_2<a_3\to+\infty)$.

Further inspection to Fig.\,\ref{f:sani}
shows additional features, namely: (i)
null left first derivatives on each
sequence at bifurcation point (not
marked therein for sake of clarity), and (ii)
occurrence of symmetric sequences with
respect to the horizontal axis (including
also forbidden configurations).

The existence of null left first
derivatives on each sequence, at
bifurcation point, cannot be deduced
analytically but, on the other hand,
it may be inferred both numerically
and graphycally.   Further support
to this idea comes from the following
considerations.   Sequences branching
from the bifurcation point on the left,
such as those plotted in Fig.\,\ref
{f:sani}, are related to triaxial
configurations where the equatorial
axis ratio does not exceed unity,
$\epsilon_{21}\le1$.   The above
mentioned sequence can also be
extended to adjoint configurations
where $\epsilon_{21}\to1$, $\epsilon_
{21}\ge1$, which corresponds to
replacing each equatorial axis with
the other.   Accordingly, the extended
sequence of triaxial configurations is
symmetric with respect to the axis,
$\epsilon_{31}=(\epsilon_{31})_{bif}$
i.e. the value related to the bifurcation
point, which implies the occurrence of
an extremum point of maximum or minimum,
and then a null first derivative.
The above results may be reduced to a
single statement.
\begin{trivlist}
\item[\hspace\labelsep{\bf Theorem 8.}] \sl
Given a sequence of homeoidally striated
Jacobi ellipsoids, the contribution of
residual motion excess, $(\upsilon_N)_{ani}$,
to the rotation parameter, $(\upsilon_N)_
{iso}$, has a null left first derivative
at the bifurcation point, \linebreak
$[\diff(\upsilon_
N)_{ani}/\diff\epsilon_{31}]_{(\epsilon_
{31})_{bif}^-}=0$.
\end{trivlist}

The occurrence of symmetric sequences
with respect to the horizontal axis,
is deduced from Eq.\,(\ref{eq:u3a})
using the condition:
\begin{equation}
\label{eq:sise}
(\tilde{\zeta}_{33}^+)^{-1}-3=-
(\tilde{\zeta}_{33}^-)^{-1}+3~~;
\end{equation}
where $\tilde{\zeta}_{33}^\mp=\zeta_{33}^
\mp/\zeta$ is related to curves
characterized by negative $(\tilde{\zeta}_
{33}=\tilde{\zeta}_{33}^-\ge1/3)$ and
positive $(\tilde{\zeta}_{33}=\tilde
{\zeta}_{33}^+\le1/3)$ values,
respectively, of the rotation parameter,
$(\upsilon_N)_{ani}$, see Fig.\,\ref
{f:sani}.   Couples of symmetric
sequences start from $(\tilde{\zeta}_
{33}^-,\tilde{\zeta}_{33}^+)=(1/3,
1/3)$, where each curve coincides
with the other, and end at (1,1/5).
Sequences in the range, $0\le\tilde
{\zeta}_{33}^+<1/5$, have no symmetric
counterpart.

Let a point, ${\sf P}[\epsilon_{31},
(\upsilon_N)_{iso}]$, be fixed on
the extended sequence of classical
Jacobi ellipsoids, and the axis ratio,
$\epsilon_{21}$, be determined by use
of Eqs.\,(\ref{eq:uq3i}), (\ref
{eq:uqi}).   Let a point, ${\sf P}^
\prime[\epsilon_{31},
(\upsilon_N)_{ani}]$, be fixed on
the plane, $[{\sf O}\epsilon_{31}
(\upsilon_N)_{ani}]$, and the
generalized anisotropy parameters,
$\tilde{\zeta}_{11}$, $\tilde{\zeta}_
{22}$, $\tilde{\zeta}_{33}$, be
determined by use of
Eqs.\,(\ref{eq:uq3a}), (\ref
{eq:u3a}), (\ref{seq:z12}).   Finally,
let the normalized rotation parameter,
$\upsilon_N$, be determined by use of
Eq.\,(\ref{eq:uNia}).   Following the
above procedure, sequences of homeoidally
striated Jacobi ellipsoids may be
generated.   For assigned density
profiles and systematic rotation
velocity fields, there are three
independent parameters, which may
be chosen as two axis ratios,
$\epsilon_{21}$, $\epsilon_{31}$,
and one generalized anisotropy
parameter, $\tilde{\zeta}_{33}$.

\subsection{Centrifugal support along the
major equatorial axis}\label{cesu}

The calculation of the gravitational force,
induced by homeoidally striated Jacobi
ellipsoids, involves numerical integrations
(e.g., Chandrasekhar 1969, Chap.\,3, \S\,20)
and is outside the aim of the current paper.
With regard to an assigned isopycnic surface
internal to, or coinciding with, the
boundary, the gravitational force, $F_G$,
induced at the end of the related major
equatorial semiaxis, $a_1^\prime$, satisfies
the inequality:
\begin{displaymath}
[F_G(a_1^\prime,0,0)]_{sph}\le F_G(a_1^\prime,
0,0)\le[F_G(a_1^\prime,0,0)]_{foc}~~;
\end{displaymath}
where the left-hand side is related to
spherical isopycnics with unchanged major
equatorial axes, and the right-hand side
to confocal isopycnic surfaces from the
one under consideration to the centre.
Owing to Newton's and MacLaurin's theorem
(e.g., Caimmi 2003), the following relations
hold:
\begin{equation}
\label{eq:FG}
F_G(a_1^\prime,0,0)=-2\pi G\bar{\rho}_w
(a_1^\prime)(A_1)_wa_1^\prime~~;
\end{equation}
where $\bar{\rho}(a_1^\prime)$ is the mean
density within the isopycnic surface, and
$w=sph,\,foc,$ for the striated sphere and
the focaloidally striated ellipsoid
surrounded by the homeoidally striated
corona, respectively.

The balance between gravitational and
centrifugal force at the end of the major
equatorial axis of the isopycnic surface
under discussion, reads:
\begin{equation}
\label{eq:eqro}
-2\pi G\bar{\rho}_w(a_1^\prime)(A_1)_wa_1
^\prime+\Omega_{iso}^2(a_1^\prime)a_1^
\prime=0~~;
\end{equation}
where the angular velocity, $\Omega_
{iso}$, takes into account both
systematic rotation and residual motion
excess, according to Eq.\,(\ref{eq:O2i}).
On the other hand, the generalization
of Eq.\,(\ref{eq:vg}) to a generic
isopycnic surface within the boundary,
reads:
\begin{equation}
\label{eq:ua1p}
\upsilon(a_1^\prime)=\frac{\Omega^2(a_1^
\prime)}{2\pi G\bar{\rho}_w(a_1^\prime)}~~;
\end{equation}
and the combination of Eqs.\,(\ref{eq:upzN}),
(\ref{eq:uNo}), (\ref{eq:O2i}), (\ref
{eq:eqro}), (\ref{eq:ua1p}), yields:
\begin{equation}
\label{eq:ua1e}
\{[\upsilon_{iso}(a_1^\prime)]_{eq}\}_w
=\frac{\Omega^2_{iso}(a_1^\prime)}{2\pi G\bar
{\rho}_w(a_1^\prime)}=(A_1)_w~~;
\end{equation}
where the index, $eq$, denotes the
rotation parameter, $\upsilon_{iso}(a_1^
\prime)$, related to centrifugal
support at the end of major equatorial
axis of the isopycnic surface under
consideration.   The above results
allow the validity of the relation:
\begin{equation}
\label{eq:eqru}
\{[\upsilon_{iso}(a_1^\prime)]_{eq}\}_{cof}
\le[\upsilon_{iso}(a_1^\prime)]_{eq}\le
\{[\upsilon_{iso}(a_1^\prime)]_{eq}\}_{sph}~~;
\end{equation}
where $[\upsilon_{iso}(a_1^\prime)]_{eq}$
is the critical value related to the
homeoidally striated Jacobi ellipsoid,
with regard to an assigned isopycnic
surface internal to, or coinciding with,
the boundary.   Then
$\upsilon_{iso}(a_1^\prime)\ge\{[\upsilon_
{iso}(a_1^\prime)]_{eq}\}_{cof}$ and
$\upsilon_{iso}(a_1^\prime)\le\{[\upsilon_
{iso}(a_1^\prime)]_{eq}\}_{sph}$ make a
sufficient and a necessary condition,
respectively, for the occurrence of
centrifugal support at the end of
major equatorial axis in the case
under consideration.

Owing to Newton,s and MacLaurin's
theorem, the condition of centrifugal
support, Eq.\,(\ref{eq:ua1e}), for
focaloidally striated ellipsoids
surrounded by homeoidally striated coronae,
coincides with its counterpart related
to homogeneous ellipsoids with equal
mean density, axis ratios, and velocity
field.   On the other hand, the
rotation parameter, $\upsilon_{iso}
(a_1^\prime)$, in the special case
of focaloidally striated ellipsoids,
is also expressed by Eq.\,(\ref{eq:ua1e}),
particularized to classical Jacobi
ellipsoids.  A comparison with the
last part of Eq.\,(\ref{eq:ua1e})
shows that centrifugal support in
focaloidally striated ellipsoids
occurs only for flat configurations,
$\epsilon_{31}=0$.   Accordingly,
Eq.\,(\ref{eq:eqru}) reduces to:
\begin{equation}
\label{eq:eqes}
0\le[\upsilon_{iso}(a_1^\prime)]_{eq}\le
\frac23~~;
\end{equation}
owing to Eq.\,(\ref{eq:ua1e}) and
$(A_1)_{sph}=2/3$ (e.g., CM03).

With regard to rigid rotation, it is a
well known result that the occurrence
of centrifugal support, at the end of
the major equatorial axis, depends on
the steepness of the density profile
(e.g., Jeans 1929, Chap.\,IX,
\S\S\,230-240).   More precisely, a
steeper density profile implies an
earlier centrifugal support and vice
versa, unless the bifurcation point
from ellipdoidal to pear-shaped
configurations is attained, which
occurs for nearly homogeneous matter
distributions.   A similar trend is
expected to hold for any velocity
profile of the kind considered in
the current paper.

The combination of Eq.\,(\ref{eq:eqru})
with its counterpart related to the
boundary, yields:
\begin{equation}
\label{eq:uapa}
\upsilon_{iso}(a_1^\prime)=\upsilon_{iso}
\frac M{M(a_1^\prime)}\left(\frac{a_1^\prime}
{a_1}\right)^3\frac{\Omega^2(a_1^\prime)}
{\Omega^2(a_1)}~~;
\end{equation}
a further restriction to mass distributions
obeyng the following law:
\begin{equation}
\label{eq:Mak}
\frac{M(a_1^\prime)}M=\left(\frac{a_1^\prime}
{a_1}\right)^k~~;\qquad0\le k\le3~~;
\end{equation}
where $k=3,1,0,$ represent homogeneous,
isothermal, and Roche-like mass
distributions, make Eq.\,(\ref{eq:uapa})
reduce to:
\begin{equation}
\label{eq:uMak}
\upsilon_{iso}(a_1^\prime)=\upsilon_{iso}
\left(\frac{a_1^\prime}{a_1}\right)^{3-k}
\frac{\Omega^2(a_1^\prime)}{\Omega^2(a_1)}~~;
\end{equation}
where the last factor equals unity for
rigid rotation and the ratio, $(a_1/a_1
^\prime)^2$, for constant velocity along
the major equatorial axis.   Accordingly,
centrifugal support at the end of major
equatorial axis is first attained on the
boundary in the former alternative,
$\upsilon_{iso}(a_1^\prime)\le\upsilon_
{iso}(a_1)$.

In the latter alternative, Eq.\,(\ref
{eq:uMak}) reduces to:
\begin{equation}
\label{eq:uMac}
\upsilon_{iso}(a_1^\prime)=\upsilon_{iso}
\left(\frac{a_1^\prime}{a_1}\right)^{1-k}~~;
\end{equation}
which, for sufficiently steep
density profiles, $0\le k\le1$,
shows a similar trend with respect
to the former alternative.
On the other hand, centrifugal
support is first attained everywhere
along the major equatorial axis for
isothermal mass distributions, $k=1$,
and at the centre for sufficiently
mild density profiles, $1\le k\le3$,
which implies $\upsilon_{iso}(a_1^
\prime)\ge\upsilon_{iso}(a_1)$.

The combination of Eqs.\,(\ref{eq:upzN})
and (\ref{eq:ua1e}) yields:
\begin{equation}
\label{eq:uNip}
\{[(\upsilon_N)_{iso}(a_1^\prime)]_{eq}\}_w=
\frac{3\eta_{rot}\nu_{rot}}{\nu_{sel}}(A_1)_w~~;
\end{equation}
and Eq.\,(\ref{eq:eqes}) reads:
\begin{equation}
\label{eq:uNie}
0\le[(\upsilon_N)_{iso}(a_1^\prime)]_{eq}\le
\frac{2\eta_{rot}\nu_{rot}}{\nu_{sel}}~~;
\end{equation}
in terms of the normalized rotation
parameter, $\upsilon_N$.

\subsection{Bifurcation points}
\label{bipo}

Owing to Theorem 4, a bifurcation
point from axisymmetric to triaxial
configurations is related to an axis
ratio, $\epsilon_{31}$, which is
independent of the amount of systematic
rotation and residual motion excess,
according
to Eq.\,(\ref{eq:uNo}).   To gain more
insight, let us equalize the alternative
expressions of Eq.\,(\ref{eq:uNq3}).
The result is:
\begin{equation}
\label{eq:Az13}
A_1-\frac{\zeta_{11}}{\zeta_{33}}\epsilon_
{31}^2A_3=A_2-\frac{\zeta_{22}}{\zeta_{33}}
\epsilon_{32}^2A_3~~;
\end{equation}
and the combination of Eqs.\,(\ref{seq:z12})
and (\ref{eq:Az13}) yields:
\begin{equation}
\label{eq:epA}
\frac{A_2-A_1}{1-\epsilon_{21}^2}=\frac
{\epsilon_{31}^2}{\epsilon_{21}^2}A_3~~;
\end{equation}
and the bifurcation points occur at a
configuration, where the axis ratio,
$\epsilon_{31}$, is the solution of
the transcendental equation (Caimmi
1996a):
\begin{equation}
\label{eq:bif1}
\lim_{\epsilon_{21}\to1}\epsilon_{21}^2
\frac{A_2-A_1}{1-\epsilon_{21}^2}=
\epsilon_{31}^2A_3~~;
\end{equation}
where Eqs.\,(\ref{eq:epA}) and (\ref
{eq:bif1}) coincide with their
counterparts related to isotropic
residual velocity distributions
(Caimmi 1996a,b).   The above results
may be reduced to a single statement.
\begin{trivlist}
\item[\hspace\labelsep{\bf Theorem 9.}] \sl
Given a sequence of homeoidally striated
Jacobi ellipsoids, the bifurcation point
occurs at the same configuration, as in
the sequence of classical Jacobi ellipsoids.
\end{trivlist}
The contradiction with earlier results
(Wiegandt 1982a,b; CM05) is explained
in the following way.   Let us suppose
that the generalized anisotropy parameters,
$\zeta_{11}, \zeta_{22}, \zeta_{33},$
can be arbitrarily fixed regardless 
from Eqs.\,(\ref{seq:z12}), and assume
$\zeta_{11}=\zeta_{22}$ {\it also} for
triaxial configurations.   Accordingly,
Eq.\,(\ref{eq:Az13}) reads:
\begin{equation}
\label{eq:epAz}
\frac{A_2-A_1}{1-\epsilon_{21}^2}=
\frac{\zeta_{11}}{\zeta_{33}}\frac
{\epsilon_{31}^2}{\epsilon_{21}^2}A_3~~;
\end{equation}
and the bifurcation points occur at a
configuration where the axis ratio,
$\epsilon_{31}$, is the solution of
the transcendental equation:
\begin{equation}
\label{eq:bifz}
\lim_{\epsilon_{21}\to1}\epsilon_{21}^2
\frac{A_2-A_1}{1-\epsilon_{21}^2}=
\frac{\zeta_{11}}{\zeta_{33}}
\epsilon_{31}^2A_3~~;
\end{equation}
which coincides with Wiegandt (1982b)
criterion for bifurcation, with regard
to homeoidally striated Jacobi ellipsoids
in rigid rotation.   For a formal
demonstration, see Appendix B.   Then
Wiegandt's criterion for bifurcation,
expressed by Eq.\,(\ref{eq:bifz}),
is in contradiction with Eqs.\,(\ref
{seq:z12}), contrary to the current one,
expressed by Eq.\,(\ref{eq:bif1}).

\subsection{Interpretation of results from
numerical simulations on stability}
\label{insi}

Given a homeoidally striated Jacobi
ellipsoid, the occurrence of
anisotropic residual velocity
distribution may be related to
residual motion excess, via
Eqs.\,(\ref{eq:uq3a}) and
(\ref{eq:u3a}).   In fact,
the changes:
\begin{leftsubeqnarray}
\slabel{eq:cEra}
&& (E_{res})_{qq}\to(E_{res})_{qq}-(\Delta E_
{res})_{qq}~~; \\
\slabel{eq:cErb}
&& (E_{rot})_{qq}\to(E_{rot})_{qq}+(\Delta E_
{res})_{qq}~~; \\
&& (\Delta E_{res})_{qq}=(\zeta_{33}-\zeta_
{qq})E_{res}~~;\qquad q=1,2~~;
\label{seq:cEr}
\end{leftsubeqnarray}
do convert (positive or negative) peculiar
motion excess into systematic
(real or imaginary) rms rotation, and the
diagonal components of the residual-energy
tensor read $(E_{res})_{pp}\to(E_{res})_{pp}-
(\Delta E_{res})_{pp}=(E_{res})_{33}$.

The velocity field related to rms rotation
has to coincide with the pre-existing one,
apart from a renormalization, defined by
Eq.\,(\ref{eq:O2a}), and the mean rotational
velocity, $\Omega_{ani}(x_1,x_2,x_3)(x_1^2+
x_2^2)^{1/2}$, has necessarily to be null
within any infinitesimal mass element.

In this view, recent results from numerical
simulations on the stability of rotating
stellar systems (Meza 2002), find a natural
interpretation.   Each computer run therein
proceeds along the following steps. 
\begin{description}
\item[\rm{(i)}\hspace{0.2mm}] Take a
spherical system with isotropic peculiar
velocity distribution and null systematic
rotation velocity field.
\item[\rm{(ii)}~~] Fix a rotation axis,
let it be $x_3$.
\item[\rm{(iii)}~] Reverse a selected sense
of rotation of a given fraction of the
particles, by changing the tangential
(parallel to equatorial plane) velocity 
component, $v_\phi$, into its opposite,
$-v_\phi$.
\item[\rm{(iv)}\hspace{1.7mm}] Make the 
system evolve from the above defined
initial configuration.
\end{description}
It is worth noticing (Meza 2002) that
the distribution function is independent
of the sign of $v_\phi$, and the whole
set of possible initial configurations
under discussion, are characterized by
the same total kinetic and potential
energy.

The simulations show that spherical
systems of the kind considered are
dynamically stable, irrispective of
their degree of rotation.   Even if
very rapidly rotating, boundaries
fail to become oblate-like.

In the light of the current theory,
reversing an assigned sense of
rotation of a given fraction of
the particles, implies a conversion
of a fraction of residual energy
into rotational energy, as:
\begin{lefteqnarray}
\label{eq:DE}
&& \Delta E_{res}=-\Delta E_{rot}
=-nm<v_\phi^2>~~; \\
\label{eq:mvfi}
&& <v_\phi^2>=\frac1n\sum_{i=1}^n\left[
v_\phi^{(i)}\right]^2~~;\qquad0\le n\le
\frac N2~~; \\
\label{eq:svf0}
&& \sum_{i=1}^{N-2n}v_\phi^{(i)}=0~~;
\qquad n\ll\frac N2~~;
\end{lefteqnarray}
where $n$ is the number of particles with
reversed tangential (parallel to the
equatorial plane) velocity component,
$N$ is the total number of particles,
and $m$ is the mass, assumed to be equal
for each particle, as done in numerical
simulations (Meza 2002).   The validity
of Eq.\,(\ref{eq:svf0}) does not necessarily
rule out streaming motions other than
systematic rotation.   On the contrary,
streaming motions are allowed provided
the existence of a given flow implies a
counter-flow, such that Eq.\,(\ref
{eq:svf0}) is satisfied.

Before reverting the sense of rotation,
the following relations hold in consequence
of isotropic residual velocity distribution:
\begin{leftsubeqnarray}
\slabel{eq:rmsia}
&& <v_1^2>=<v_2^2>=<v_3^2>~~; \\
\slabel{eq:rmsib}
&& <v_\phi^2>+<v_w^2>=<v_1^2>+<v_2^2>~~;
\label{seq:rmsi}
\end{leftsubeqnarray}
where $v_p$, $p=1,2,3$, and $v_w$, $w=(x_1^2+
x_2^2)^{1/2}$, are
velocity components parallel to coordinate
axes and equatorial plane, respectively.
After reversion of the sense of rotation,
the following changes are performed:
\begin{lefteqnarray}
\label{eq:cres}
&& E_{res}\to E_{res}-nm<v_\phi^2>~~; \\
\label{eq:cref}
&& (E_{res})_\phi\to(E_{res})_\phi-
nm<v_\phi^2>~~; \\
\label{eq:creq}
&& (E_{res})_{qq}\to(E_{res})_{qq}-
\frac12nm<v_\phi^2>~~;\qquad q=1,2~~; \\
\label{eq:crot}
&& E_{rot}\to0+nm<v_\phi^2>~~; \\
\label{eq:croq}
&& (E_{rot})_{qq}\to0+\frac12nm
<v_\phi^2>~~;\qquad q=1,2~~;
\end{lefteqnarray}
while the contributions from residual
motions along the equatorial plane,
$(E_{res})_w$, and the rotation axis,
$(E_{res})_{33}$, remain unchanged.

The system being relaxed, $\zeta=1$,
in the case under discussion, the
generalized and effective anisotropy
parameters, $\zeta_{pp}$ and $\tilde
{\zeta}_{pp}$, coincide with their
usual counterparts, and Eqs.\,(\ref
{eq:zepq}) and (\ref{eq:tzpq}) take
the explicit form:
\begin{leftsubeqnarray}
\slabel{eq:zvfa}
&& \zeta_{qq}=\frac{(E_{res})_{qq}-
(n/2)m<v_\phi^2>}{(E_{res})-nm<v_
\phi^2>}~~;\qquad q=1,2~~; \\
\slabel{eq:zvfb}
&& \zeta_{33}=\frac{(E_{res})_{33}}
{(E_{res})-nm<v_\phi^2>}~~;
\label{seq:zvf}
\end{leftsubeqnarray}
where the special case, $n=0$, turns 
back to isotropic residual velocity
distributions, $\zeta_{11}=\zeta_{22}
=\zeta_{33}=1/3$.

In the extreme case where the sense
of rotation is completely reversed,
$n=N/2$, the changes expressed by
Eqs.\,(\ref{eq:cres})-(\ref{eq:croq}) 
reduce to:
\begin{lefteqnarray}
\label{eq:EreN}
&& E_{res}\to E_{res}-\frac N2m
<v_\phi^2>=\frac23E_{res}~~; \\
\label{eq:ErfN}
&& (E_{res})_\phi\to(E_{res})_\phi-
\frac N2m<v_\phi^2>=0~~; \\
\label{eq:ErqN}
&& (E_{res})_{qq}\to(E_{res})_{qq}-
\frac N4m<v_\phi^2>=\frac16E_{res}~~;
\qquad q=1,2~~; \\
\label{eq:EroN}
&& E_{rot}\to0+\frac N2m<v_\phi^2>=
\frac13E_{res}~~; \\
\label{eq:Eoqn}
&& (E_{rot})_{qq}\to0+\frac N4m
<v_\phi^2>=\frac16E_{res}~~;\qquad
q=1,2~~;
\end{lefteqnarray}
and the anisotropy parameters,
expressed by Eqs.\,(\ref{seq:zvf}),
reduce to:
\begin{leftsubeqnarray}
\slabel{eq:zvNa}
&& \zeta_{qq}=\frac{(1/6)E_{res}}
{(2/3)E_{res}}=\frac14~~;\qquad q=1,2~~; \\
\slabel{eq:zvNb}
&& \zeta_{33}=\frac{(1/3)E_{res}}
{(2/3)E_{res}}=\frac12~~;
\label{seq:zvN}
\end{leftsubeqnarray}
which describes the maximum amount
of residual motion excess, related
to complete reversion of the sense
of the rotation.

Turning back to the general case,
a comparison between the
expression of rotational energy,
Eq.\,(\ref{eq:Er}), with the special
one related to the case under discussion,
Eq.\,(\ref{eq:crot}), reads:
\begin{equation}
\label{eq:gpn}
\eta_{rot}\nu_{rot}M(1+\epsilon_{21}^2)
<v_\phi^2>_{a_1}=nm<v_\phi^2>~~;
\end{equation}
where $<v_\phi^2>_{a_1}$ is the rms
systematic rotation velocity at the
end of the major equatorial \linebreak
axis
\footnote{With the additional restriction,
that streaming motions within the volume
element at the end of major equatorial
axis are allowed provided the existence
of a given flow therein implies a
counter-flow, such that Eq.\,(\ref
{eq:svf0}) is satisfied in the region
of interest.}, %
$w=(x_1^2+x_2^2)^{1/2}=a_1$.   The
related rms angular velocity is:
\begin{equation}
\label{eq:O2f}
\Omega^2=\Omega^2(a_1)=\frac
{<v_\phi^2>_{a_1}}{a_1^2}~~;
\end{equation}
and the combination of Eqs.\,(\ref
{eq:Es}), (\ref{eq:Bspq}),
(\ref{eq:virte}), (\ref{eq:vg}),
(\ref{eq:rome}), (\ref{eq:upzN}),
(\ref{eq:svf0}), (\ref{eq:gpn}), and
(\ref{eq:O2f}) yields after some
algebra:
\begin{equation}
\label{eq:uNf}
\upsilon_N=\frac{2nm<v_\phi^2>a_1
\epsilon_{21}\epsilon_{31}}{\nu_
{sel}GM^2(1+\epsilon_{21}^2)}=-
\frac{B_{sel}\epsilon_{21}\epsilon_
{31}}{1+\epsilon_{21}^2}\frac{\Delta
E_{res}}E~~;
\end{equation}
where $E=E_{sel}/2$ is the total
energy, according to the virial
theorem.   On the other hand, the
combination of Eqs.\,(\ref{eq:u3a}),
(\ref{eq:svf0}), and (\ref{seq:zvf})
yields:
\begin{equation}
\label{eq:ufa}
(\upsilon_N)_{ani}=-\frac{3nm<v_\phi^2>}
{E_{res}}\frac{\epsilon_{31}^2A_3}{1+
\epsilon_{21}^2}=-3\frac{\epsilon_{31}^2
A_3}{1+\epsilon_{21}^2}\frac{\Delta E_
{res}}E~~;
\end{equation}
where $E=-E_{res}$ is the total
energy, according to the virial
theorem.

The validity of the above results
is related to the assumption of
relaxed configurations, which
implies a virial index, $\zeta=1$,
according to Eqs.\,(\ref{eq:virte}),
(\ref{eq:zepq}), and (\ref{eq:zeta}).
In addition, the assumption of
sphericity necessarily restricts
the validity of Eqs.\,(\ref{eq:uNf})
and (\ref{eq:ufa}) to spherical
systems, where $\epsilon_{21}=
\epsilon_{31}=1$, $A_1=A_2=A_3=
2/3$, $B_{sel}=2$.   Accordingly,
the expression of the rotation
parameters, $\upsilon_N$ and
$(\upsilon_N)_{ani}$, reduces to:
\begin{equation}
\label{eq:ufs}
\upsilon_N=\frac{\Delta E_{res}}E~~;\qquad
(\upsilon_N)_{ani}=-\frac{\Delta E_{res}}E~~;
\end{equation}
which implies an adjoint spherical
configuration with rotation parameter,
$(\upsilon_N)_{iso}=\upsilon_N+
(\upsilon_N)_{ani}=0$, as expected.
This is why an additional real
rotation, due to the change $E_{rot}
\to E_{rot}-\Delta E_{res}$, is exactly
balanced by an additional imaginary
rotation, due to the change $E_{res}
\to E_{res}+\Delta E_{res}$, according to
Eqs.\,(\ref{eq:ufs}).   Then the
initial shape remains unchanged.
In addition, owing to Theorem 4,
the condition of stability is also
expected to be left unaltered.
In other words, the rotating sphere
$(\zeta_{11}=\zeta_{22}<\zeta_{33})$
is expected to be as stable as the
nonrotating sphere
$(\zeta_{11}=\zeta_{22}=\zeta_{33})$,
regardless from the assumed density
profile, according to the results
from numerical simulations (Meza 2002).

\section{Application: physical interpretation
of the early Hubble sequence}
\label{elli}

The results of the current paper allow
an extension and a generalization to
anisotropic peculiar velocity
distributions, of the classical physical
interpretation of the Hubble (1926)
sequence (e.g., Jeans 1929, Chap.\,XIII,
\S\S\,298-303).   In the following,
quotations are from therein.   Owing
to Theorem 4, it will suffice to
restrict to solid-body rotating
configurations with isotropic peculiar
velocity distributions, but with
imaginary rotation also taken into
consideration.

\subsection{Classification of galaxies}
\label{clag}

It is worth recalling that, before
establishing their similarity to the
Milky Way, galaxies were referred to
as ``great nebulae'' or ``nebulae''.
\begin{quotation}
Hubble finds that it is not possible
to place all observed nebulae in one
continuous sequence; their proper
representation demands a ${\sf Y}$-shaped
diagram. (...)

The lower half of the ${\sf Y}$ is
formed by nebulae of approximately
elliptical or circular shape.
These are subdivided into eight
classes, designated $E0$, $E1$, ...,
$E7$, the numerical index being the
integer nearest to $10[(a-b)/a]$,
where $a$ and $b$ are the greatest
and the least diameter of the
nebulae as projected on the sky.
Thus $E0$ consists of nearly
circular nebuale $(b>0.95a)$,
while $E7$ consists of nebulae
for which $b$ is about $0.3a$,
this being the greatest inequality
of axes observed in the ``elliptical''
nebulae. (...)

The upper half of the ${\sf Y}$-shaped
diagram consists of two distinct
branches, one of which is found to
contain a far larger number of
nebulae than in the other.   The
principal branch contains the normal
``spiral'' nebulae, which are
characterized by a circular nucleus
from which emerge two (or occasionally
more) arms of approximately spiral
shape. (...)   These nebulae are
subdivided into
three classes, designated $Sa$, $Sb$,
$Sc$, class $Sa$ fitting almost
continuously on to class $E7$.

The minor branch contains a special
class of spirals characterized by
the circumstance that the spiral arms
appear to emerge from the two ends of
a straight bar-shaped or spindle-shaped
mass. (...)

About 97 per cent of known extra-galactic
nebulae are found to fit into this ${\sf
Y}$-shaped classification.   The remaining
3 per cent, are of irregular shape, and
refuse to fit into the classification at
all. (...)   The irregular nebulae are
distinguished by a complete absence of
symmetry of figure and also by the
absence of any central nucleus.   (...)

Apart from the irregular nebulae, Hubble
states that, out of more than a thousand
nebulae examined, less than a dozen refused
to fit into the ${\sf Y}$-shaped diagram
at all, while in less than 10 per cent of
the cases was there any considerable doubt
as to a proper position of a nebula in the
diagram.   Clearly, the ${\sf Y}$-shaped
diagram provides a highly satisfactory
working classification.
\end{quotation}

Thought some improvements have been made
to the classification of galaxies (e.g.,
van den Bergh 1960a,b; Roberts \& Haynes
1994), still
we shall maintain Hubble (1926) diagram
as a valid point of reference,
owing to its intrinsic simplicity.

\subsection{Physical interpretation}
\label{fisi}

\begin{quotation}
Obviously the proper physical interpretation
of the classification just described is of
the utmost importance to cosmogony.

A first and most important clue is provided
by the fact that numbers of the great
nebulae are known to be in rotation. (...)

The symmetry of figure shewn by nebulae
of the $E$ and $Sa$ types is precisely 
such as rotation might be expected to
produce, and this suggests an inquiry
as to how far the observed figures of
the regular nebulae can be explained
as the figures assumed by masses rotating
under their own gravitation. (...)

Remembering that rotation has actually
been observed in a number of nebulae,
there seem to be strong reasons for
conjecturing that the observed configuration
of the nebulae may be explained in general
terms as the configurations of rotating
masses.
\end{quotation}

More specifically, two classes of models
are considered, where density profiles
range between the limiting cases (i)
homogeneous configurations
i.e. classical MacLaurin spheroids and
Jacobi ellipsoids, and (ii) mass points
surrounded by massless atmospheres i.e.
Roche systems.   Rigidly rotating mass
distributions of the kind considered
may attain any shape between the extreme
boundaries (a) spherical i.e. nonrotating,
and (b) centrifugal support at the end
of major equatorial axis.   With respect
to the latter configuration, further
rotation implies equatorial shedding
(sequence SN) or top major equatorial
axis streaming (sequenceSB) of matter,
for axisymmetric and triaxial configurations,
respectively.   For a fixed density profile,
the sequence of rigidly rotating
configurations starts from the spherical
shape and ends at the shape where centrifugal
support is first attained.

The above interpretation suffers from two
main points.   First, the most flattened
elliptical configuration should occur when
centrifugal support takes place at the
bifurcation point, from axisymmetric to
triaxial shapes.   On the other hand, the
above mentioned limiting configuration is
less flattened than $E7$.   Second,
triaxial bodies of the kind considered
cannot be figures of equilibrium if a
great central condensation of mass is
present.   On the other hand, it is the
case for barred spirals.

Owing to Theorem 4, homeoidally
striated Jacobi ellipsoids with arbitrary
peculiar velocity field, and systematic
motions reduced to rotation around a
fixed axis, may be related to classical
Jacobi ellipsoids, provided imaginary
rotation is taken into consideration.
Then the occurrence of anisotropic
peculiar velocity distributions does
not affect the physical interpretation
under discussion, and the questions
to be clarified are the above mentioned
two.

\subsection{The $E$ sequence within
Ellipsoidland}
\label{Esel}

In dealing with a physical interpretation
of the early Hubble sequence, it is convenient
to define axis ratios of intrinsic
configurations in a different way.
Given a homeoidally striated Jacobi
ellipsoid, let $a$, $b$, $c$, be the
semiaxes, where $a\ge b\ge c$ without
loss of generality.   Accordingly,
$\epsilon_{ca}\le\epsilon_{ba}\le1$;
in addition, the minor and the
major axis coincide with the rotation
axis for oblate-like and prolate-like
configurations, respectively.

\begin{figure}
\centerline{\psfig{file=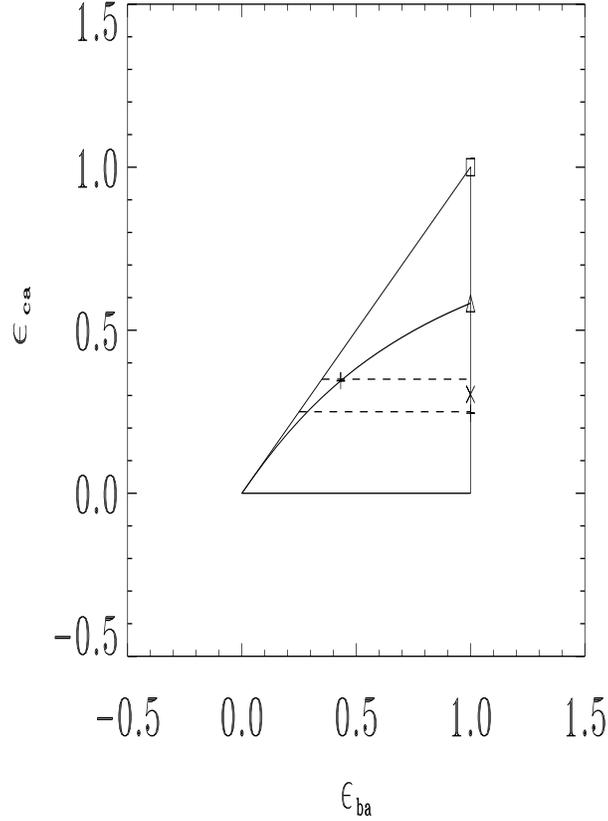,height=130mm,width=90mm}}
\caption{Axis ratio correlation within
Ellipsoidland, with regard to adjoint
configurations to homeoidally striated
Jacobi ellipsoids.   The semiaxes are
$a$, $b$, $c$, where $a\ge b\ge c$ or
$\epsilon_{ca}\le\epsilon_{ba}\le1$,
without loss of generality.   The
locus of flat configurations is
$\epsilon_{ca}=0$, $0\le\epsilon_{ba}
\le1$.   The locus of oblate configurations
is $\epsilon_{ba}=1$, $0\le\epsilon_{ca}
\le1$.   The spherical configuration is
defined as $\epsilon_{ca}=\epsilon_{ba}
=1$.   The oblong configuration is defined
as $\epsilon_{ca}=\epsilon_{ba}=0$.
The locus of prolate configurations is
$\epsilon_{ca}=\epsilon_{ba}$, $0\le
\epsilon_{ba}\le1$.   Different
configurations occur in the flat oblong
limit, $\epsilon_{ca}=0$, $\epsilon_{ba}
\to0$, and in the prolate oblong limit,
$\epsilon_{ca}\to0$, $\epsilon_{ba}\to0$,
$\epsilon_{ca}=\epsilon_{ba}$.  A band,
bounded by two horizontal dashed lines,
defines class $E7$ in
Hubble (1926) classification i.e. $0.25
<\epsilon_{ca}\le0.35$ for edge-on
ellipsoids where the line of sight
coincides with the direction of the
minor equatorial axis.   Other symbols
as in Fig.\,\ref{f:siso}.}
\label{f:elli}
\end{figure}

The whole range of possible configurations
in the $({\sf O}\epsilon_{ba}\epsilon_
{ca})$ plane defines Ellipsoidland (the
term is from Hunter \& de Zeeuw 1997).
Ellipsoidland is a triangle where
two sides are of unit length and an
angle is right, as shown in Fig.\,\ref
{f:elli}.   Oblate configurations lie
on the vertical side, $\epsilon_{ba}=1$,
the spherical configuration is on the top,
$\epsilon_{ca}=1$, and the
flat circular configuration is on the
bottom, $\epsilon_{ca}=0$.   Prolate
configurations lie on the inclined
side, $\epsilon_{ca}=\epsilon_{ba}$,
the spherical configuration is on the top,
$\epsilon_{ca}=\epsilon_{ba}=1$, and the
prolate oblong configuration is on the
bottom, $\epsilon_{ca}=\epsilon_{ba}=0$.
Flat configurations lie on the horizontal
side, $\epsilon_{ca}=0$, the 
flat circular configuration is on the
right, $\epsilon_{ba}=1$, and the
flat oblong configuration is on the
left, $\epsilon_{ba}=0$.  It is worth
noticing that flat-oblong and
prolate-oblong configurations are not
coincident (e.g., Caimmi 1993; CM05).
Non flat, non axisymmetric configurations,
lie within Ellipsoidland.

The axis ratio correlation, 
$\epsilon_{ca}$ vs. $\epsilon_{ba}$,
related to adjoint configurations to
homeoidally striated Jacobi ellipsoids,
is represented within Ellipsoidland
in Fig.\,\ref{f:elli}.   The caption
of symbols is the same as in Fig.\,\ref 
{f:siso}.   With regard to real rotation, 
the correlation is the known one
involving classical MacLaurin spheroids
and Jacobi ellipsoids.   With regard to
imaginary rotation, the correlation reads
$\epsilon_{ca}=\epsilon_{ba}$, owing to
lack of bifurcation points.   Then
configurations in real and imaginary
rotation branch off from the nonrotating
spherical configuration.   The former
sequence proceeds down along the oblate
side of Ellipsoidland, until the
bifurcation point is attained and
the curve enters Ellipsoidland,
finishing when the next bifurcation
point is also attained.   The latter
sequence proceeds down along the
prolate side of Ellipsoidland, until
the prolate-oblong configuration is
attained, never entering Ellipsoidland.
On the other hand, a sequence may stop
earlier, when centrifugal support at
the ends of major equatorial axis is
attained, which depends on the density
profile.   In fact, it makes the sole
difference between homeoidally striated
Jacobi ellipsoids and their adjoint
counterparts (for further details, see
CM05).

With regard to Ellipsoidland, the
observed lack of elliptical galaxies
more flattened than $E7$ translates
into the inequality, $\epsilon_{ca}
\ge0.25$, the treshold being
represented by the lower dashed
horizontal line in Fig.\,\ref{f:elli}.
The locus of edge-on configurations,
viewed along a direction coinciding
with equatorial minor axis, where
$0.25\le\epsilon_{ca}\le0.35$, and
then belonging to class $E7$, is
represented in Fig.\,\ref{f:elli}
as a band bounded by two dashed
lines, $\epsilon_{ca}=0.25$ and
$\epsilon_{ca}=0.35$, respectively.
A change in direction of the line
of sight could project an intrinsic
ellipsoid within the above mentioned
band, on a class $Ei$, $i<7$.

An inspection of Fig.\,\ref{f:elli}
shows that the bifurcation points
from triaxial and axisymmetric
configurations, towards pear-shaped
configurations, are consistent, or
marginally consistent, with the $E7$
band on Ellipsoidland.   The above
considerations provide a physical
interpretation to the observed
absence of elliptical galaxies
(and spiral bulges) more flattened
than $E7$, with regard to the
oblate-like branch of the sequence.
On the other hand, the above
interpretation cannot apply to
the prolate branch of the sequence,
as no bifurcation point has been
found therein, and some kind of
instability must be considered.

The absence of elliptical galaxies
more flattened or elongated than
$E7$ might be due to bending
instabilities, as suggested long
time ago from analytical considerations
involving homogeneous (oblate and
prolate) spheroids
(Polyachenko \& Shukhman 1979;
Fridman \& Polyachenko 1984, Vol.\,1,
Chap.\,4, Sect.\,3.3, see also
pp.\,313-322; Vol.\,2,
p.\,159) and numerical simulations
involving inhomogeneous (oblate
and prolate) spheroids
(Merritt \& Hernquist 1991; Merritt
\& Sellwood 1994).   The amount of
figure rotation seems to be unimportant
to this respect (Raha et al. 1991;
Merritt \& Sellwood 1994).

This
conclusion is supported by recent
results from $N$-body numerical
simulations, where spherically
symmetric, unstable, radially
anisotropic, one-component
$\gamma$ models were taken as
starting configurations (Nipoti
et al. 2002).   The related
end-products, in accordance with
previous results (e.g., Merrit
\& Aguilar 1985; Stiavelli \&
Sparke 1991), were found to be
in general prolate systems less
flattened than $E7$ (Nipoti et
al. 2002, Fig.\,1 therein).

The theoretical explanation
of the result, in terms of a dynamical
bending instability (Merritt \& Sellwood
1994), is generally recognized to
explain also the maximum elongation
of simulated nonbaryonic dark matter
haloes (e.g., Bett et al. 2007).

\subsection{Tidal effects from
hosting dark matter haloes}
\label{tehh}

Current $\Lambda$CDM cosmologies,
which provide a satisfactory fit
to data from primordial nucleosynthesis
and cosmic background radiation,
predict galaxies are embedded within
dark matter haloes.   Then it is a
natural question to what extent the
presence of hosting dark matter haloes
may affect the above interpretation of
the early Hubble sequence.   To this aim,
an idealized situation shall be analysed.

Let us represent elliptical galaxies
and their hosting dark haloes as 
concentric and coaxial classical
Jacobi ellipsoids, one completely
lying within the other.   Accordingly,
the two bodies must necessarily
rotate at the same extent, and/or
one of them (the outer in the case
under discussion) has to be axisymmetric.
It can be seen that the effect of
the outer ellipsoid on the inner one
is to shift bifurcation points
towards more flattened configurations
with respect to a massless embedding
subsystem (Durisen 1978; Pacheco et al.
1986; Caimmi 1996a).

More specifically, the axis ratio of
the configuration at the bifurcation
point is the solution of the transcendental
equation (Caimmi 1996a):
\begin{leftsubeqnarray}
\slabel{eq:bif2a}
&& \frac{[(\epsilon_i)_{31}]^2(A_i)_3}
{(A_i)_1}=\left[\frac{5-4[(\epsilon_i)
_{31}]^2}{3-2[(\epsilon_i)_{31}]^2}
+\frac m{y_1y_2y_3}\frac{4-4[(
\epsilon_i)_{31}]^2}{3-2[(\epsilon_i)
_{31}]^2}\frac{(A_j)_3}{(A_i)_3}\right]
^{-1}~~; \\
\slabel{eq:bif2b}
&& m=\frac{M_j}{M_i}~~;\qquad y_k=\frac
{(a_j)_k}{(a_i)_k}~~;\qquad\frac m{y_1
y_2y_3}=\frac{\rho_j}{\rho_i}~~;
\label{seq:bif2}
\end{leftsubeqnarray}
where the indices, $i$, $j$, denote
inner and outer ellipsoid, respectively.

The effect of the embedding subsystem
on the shape of the embedded configuration
at the bifurcation point, is governed
by the parameter:
\begin{equation}
\label{eq:kag}
\kappa=\frac m{y_1y_2y_3}\frac{(A_j)_3}
{(A_i)_3}~~;\qquad0\le\kappa<+\infty~~;
\end{equation}
in the special case of massless outer
ellipsoid, $\kappa=0$, Eq.\,(\ref
{eq:bif2a}) reduces to (\ref{eq:bifJ}).

Further analysis shows that the
embedded configuration, at the
bifurcation point, is as flattened
as $E7$ when the parameter, $\kappa$,
is close to unity.   Accordingly,
the following relation holds:
\begin{equation}
\label{eq:bE7}
\frac m{y_1y_2y_3}\frac{(A_j)_3}
{(A_i)_3}=1~~;\qquad0.25<(\epsilon_i)
_{31}\le0.35~~;
\end{equation}
where the shape factor ratio, $(A_j)_3
/(A_i)_3$, exceeds unity if the outer
ellipsoid is more flattened than the
inner one, and vice versa.

The generalization of Eq.\,(\ref
{eq:bE7}) to homeoidally striated
Jacobi ellipsoids demands to restart
from a generalized formulation of
the virial theorem, where the tidal
potential is also included (Brosche
et al. 1983; Caimmi et al. 1984;
Caimmi \& Secco 1992).   The
repetition of the same procedure
used in the current paper, yields:
\begin{equation}
\label{eq:b2E7}
\frac m{y_1y_2y_3}\frac{(\nu_{ij})_
{tid}}{(\nu_{i})_{sel}}\frac{(A_j)_3}
{(A_i)_3}=1~~;\qquad0.25<(\epsilon_i)
_{31}\le0.35~~;
\end{equation}
where $(\nu_{ij})_{tid}$ is an
additional factor in the expression
of the potential tidal energy,
related to the tidal action of
the embedding subsystem on the
embedded one (Caimmi 2003; CM05).

It may safely be expected that the
external boundary is less flattened
than the internal one and more 
flattened than a sphere.   Accordingly,
the shape factor ratio appearing in
Eqs.\,(\ref{eq:bE7}), (\ref{eq:b2E7}),
ranges as $1/2<(A_j)_3/(A_i)_3<1$ for
oblate configurations, and $1<(A_j)_3/
(A_i)_3<10/3$ for prolate configurations.
Then, to a first extent, $(A_j)_3/(A_i)_
3\approx1$, i.e. the boundaries are
similar and similarly placed ellipsoids.
With this restriction, the factor, $(\nu
_{ij})_{tid}$, is a profile factor,
which may be expressed by a simple
formula (Caimmi 2003):
\begin{equation}
\label{eq:nuij}
(\nu_{ij})_{tid}=-\frac98\frac{m\Xi_i}
{(\nu_i)_{mas}(\nu_j)_{mas}}w^{(ext)}
\frac{\Xi_i}{y_0}~~;
\end{equation}
where $y_0$ is the scaling radius
ratio of outer to inner subsystem,
and $w^{(ext)}(\eta)$ is an additional
profile factor.

Typical elliptical galaxies and
their hosting dark haloes may
safely be represented as homeoidally
striated Jacobi ellipsoids, where
the star and dark subsystem are
described by generalized power-law
density profiles of the kind:
\begin{leftsubeqnarray}
\slabel{eq:GPLE2a}
&& f(\xi_u)=\frac{2^\chi}{\xi_u^\gamma
(1+\xi_u^\alpha)}~~;\qquad\chi=\frac
{\beta-\gamma}\alpha~~;\qquad u=i,j~~; \\
\slabel{eq:GPLE2b}
&& \xi_u=\frac{r_u}{(r_0)_u}~~;\qquad
\Xi_u=\frac{R_u}{(r_0)_u}~~; \\
\slabel{eq:GPLE2c}
&& \xi_i=y_0\xi_j~~;\qquad y\Xi_i=y_0\Xi_j
~~; \\
\slabel{eq:GPLE2d}
&& y_0=\frac{(r_0)_j}{(r_0)_i}~~;\qquad
y=\frac{R_j}{R_i}~~;
\label{seq:GPLE2}
\end{leftsubeqnarray}
according to Eqs.\,(\ref{seq:profg}),
and the choices $(\alpha,\beta,\gamma)
=(1,4,1)$ (Hernquist 1990) and $(\alpha,
\beta,\gamma)=(1,3,1)$ (Navarro et al.
1995, 1996, 1997) are adopted to describe
the star and the dark subsystem,
respectively.   The values of input
and output parameters of the model
are listed in Tab.\,\ref{t:bif2},
with regard to a $\Lambda$CDM cosmology
where $\Omega_M=0.3$, $\Omega_\Lambda=
0.7$, $\Omega_b=0.0125h^{-2}$, $h=2^
{-1/2}$.   For further details and
references, see CM05.
\begin{table}
\begin{tabular}{llll}
\hline
\hline
\multicolumn{1}{c}{input} &
\multicolumn{1}{c}{value} &
\multicolumn{1}{c}{output} &
\multicolumn{1}{c}{value} \\
\hline
$\Xi_i$ & 40/3 & $(\nu_i)_{mas}$ & $\phantom{-}$10.38399 \\
$\Xi_j$ & \phantom{3}10 & $(\nu_j)_{mas}$ & $\phantom{-}$17.86565 \\
$(r_0)_i$/kpc & \phantom{36}3.21 & $(\nu_i)_{sel}$ &
\phantom{-1}1.44444 \\
$(r_0)_j$/kpc & \phantom{3}36.20 & $(\nu_j)_{sel}$ &
\phantom{-1}0.6268271 \\
$R_i$/kpc & \phantom{3}42.77 & $(\nu_{ij})_{tid}$ &
\phantom{-1}0.3518317 \\
$R_j$/kpc & 362.04 & $(\nu_{ji})_{tid}$ & \phantom{-1}0.2217760 \\
$M_i/10^{10}m_\odot$ & \phantom{36}8.33 & $(\nu_i)_{rot}$ &
$\phantom{-}$1/3 \\
$M_j/10^{10}m_\odot$ & \phantom{3}91.67 & $(\nu_j)_{rot}$ &
$\phantom{-}$1/3 \\
$y_0$ & \phantom{3}11.29 & $(\eta_i)_{rot}$ & $\phantom{-}$1/2 \\
$y$ & \phantom{36}8.47 & $(\eta_j)_{rot}$ & $\phantom{-}$1/2 \\
$m$ & \phantom{3}11 & $w^{(est)}$ & $-$0.3955800 \\
 & & $w^{(int)}$ & $-$3.564028 \\
 & & $(\nu_{ij})_{int}$ & \phantom{-1}1.760852 \\
 & & $(\nu_{ji})_{int}$ & \phantom{-1}1.760852 \\
 & & $\left(\frac{2\eta_{rot}\nu_{rot}}{\nu_{sel}}\right)_i$ &
 \phantom{-1}0.230769 \\
 & & $\left(\frac{2\eta_{rot}\nu_{rot}}{\nu_{sel}}\right)_j$ &
 \phantom{-1}0.531779 \\
 & & $\frac m{y^3}\frac{(\nu_{ij})_{tid}}{(\nu_i)_{sel}}$ &
 \phantom{-1}0.00441624 \\
 & & $(\epsilon_{31})_{bif}$ & \phantom{-1}0.580088 \\
\hline\hline
\end{tabular}
\caption{Values of input and output parameters in
modelling elliptical galaxies and their hosting
dark haloes as similar and similarly placed,
homeoidally striated Jacobi ellipsoids,
characterized by Hernquist (1990) and Navarro
et al. (1995, 1996, 1997) density profiles,
respectively.   The parameters of the related
$\Lambda$CDM cosmology have been chosen as
$\Omega_M=0.3$, $\Omega_\Lambda=0.7$, $\Omega_
b=0.0125h^{-2}$, $h=2^{-1/2}$.  For further
details on the output parameters see e.g.,
CM03.}
\label{t:bif2}
\end{table}

The presence of a massive halo has
little influence on the location of
the bifurcation point from axisymmetric
to triaxial configurations, which
is found to occur at an axis ratio,
$(\epsilon_{31})_{bif}=0.580088$.   A similar
result is expected to occur for the 
bifurcation point from both
axisymmetric and triaxial to
pear-shaped configurations.
Then the above mentioned points
continue to be consistent, or
marginally inconsistent, with
the $E7$ band on Ellipsoidland,
even in presence of (typical)
massive dark haloes, concerning
the oblate-like branch of the
sequence.   On the contrary, it
is suggested
the existence of some kind of
instability, which does not allow
prolate configurations in rigid
imaginary rotation, more elongated
than $E7$, even in presence of
(typical) massive dark haloes,
according to recent results from
$N$-body simulations (Nipoti el al.,
2002).

\subsection{Cosmological effects
after decoupling}
\label{cead}

About thirthy years ago, Thuan \&
Gott (1975) idealized elliptical
galaxies as MacLaurin spheroids,
resulting from virialization
after cosmological expansion
and subsequent collapse and
relaxation of their parent
density perturbation.   A
generalization of the method
to triaxial configurations and
anisotropic peculiar velocity
distributions, has been performed
in CM05, and an interested reader
is addressed therein for further
details.   Owing to Theorem 4,
the case of isotropic
peculiar velocity distribution,
involving both real and imaginary
rotation, can be considered without
loss of generality.

Our attention shall be limited to
dark matter haloes hosting giant
galaxies, as massive as about
$10^{12}m_\odot$.   Accordingly,
it is assumed $\overline{\delta}_
{rec}=0.015$ as a typical
overdensity index of the initial
configuration, taken to be at
recombination epoch, and a relaxed
final configuration i.e. $\zeta=1$.

With regard to the initial
configuration, the following
approximations hold to a good
extent: (i) spherical shape;
(ii) homogeneous mass distribution;
(iii) negligible (real and/or
imaginary) rotation energy; (iv)
negligible peculiar energy.   The
changes in mass, angular momentum,
and total energy, respectively,
during the transition from the
initial to final configuration,
are expressed by the parameters:
\begin{equation}
\label{eq:beta}
\beta_M=\frac M{M^\prime}~~;\qquad\beta_J=
\frac J{J^\prime}~~;\qquad\beta_E
=\frac E{E^\prime}~~;
\end{equation}
where the initial configuration is marked
by the prime.   For further details, see
CM05.

Though negligible with respect to the
potential and expansion energy, the
rotation energy of the initial
configuration affects the shape of
the final configuration, as:
\begin{leftsubeqnarray}
\slabel{eq:trana}
&& {\cal E}_{rot}^\prime=b\mp(b^2-c)^
{1/2}~~; \\
\slabel{eq:tranb}
&& b=\frac12(1-{\cal E}_{osc}^\prime-
{\cal E}_{pec}^\prime)~~; \\
\slabel{eq:tranc}
&& c=\beta_{\cal E}\left(\frac{{\cal S}}
{{\cal S}^\prime}\right)^2\frac{{\cal R}^\prime}
{{\cal R}}{\cal E}_{rot}\left[\frac
{2\zeta-1}{2\zeta}-\frac{\zeta-1}{\zeta}
{\cal E}_{rot}\right]~~; \\
\slabel{eq:trand}
&& \beta_{\cal E}=\frac{\beta_M^5}{\beta
_J^2\beta_E} ~~; \\
\slabel{eq:trane}
&& {\cal E}_{osc}=-\frac{E_{osc}}{E_{sel}}
~~;\qquad{\cal E}_{pec}=-\frac{E_{pec}}
{E_{sel}}~~;
\label{seq:tran}
\end{leftsubeqnarray}
where different changes in mass, angular
momentum, and tidal energy, are related
to a same final shape for an assigned
initial configuration, provided $\beta_
{\cal E}$ does not vary.   The choice,
$\beta_{\cal E}=1/3600$, in particular
$(\beta_M,\beta_J,\beta_E)=(1,60,1)$,
holds for dark matter haloes hosting
giant elliptical galaxies, and it shall
be assumed here.   For further details,
see CM05.

\begin{figure}
\centerline{\psfig{file=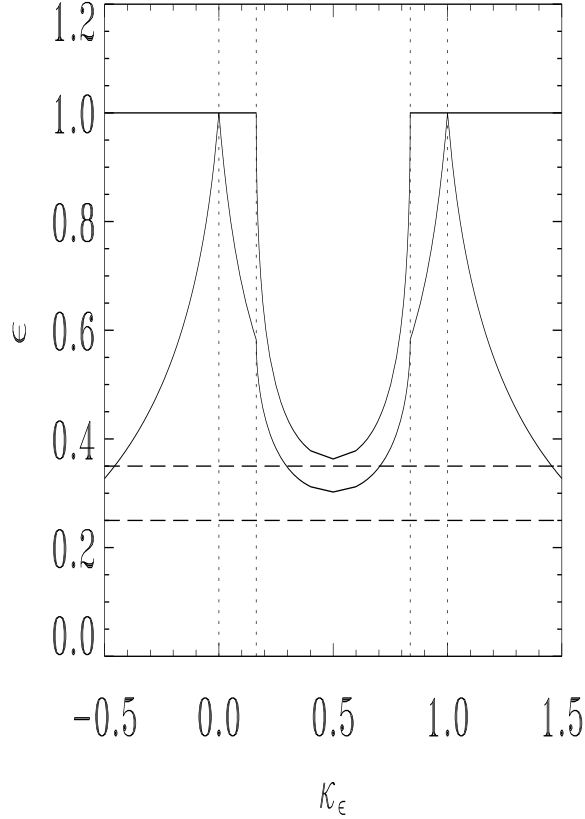,height=130mm,width=90mm}}
\caption{Equatorial (upper curve) and meridional
(lower curve) axis ratio of the final configuration,
as a function of the parameter, $\kappa_{\cal E}=
{\cal E}_{rot}^\prime/(1-{\cal E}_{osc}^\prime-
{\cal E}_{pec}^\prime)$, for sequences of relaxed
Navarro et al. (1995, 1996, 1997) density profiles
in rigid rotation, related
to dark matter haloes hosting giant elliptical
galaxies.   The curves are symmetric with respect
to a vertical axis, $\kappa_{\cal E}=0.5$, and a
minimum extremum point occurs at $(\kappa_{\cal E},
\epsilon_{21},\epsilon_{31})=(0.5, 0.363174, 0.302297)$,
corresponding to the most flattened, oblate-like
configuration
which is allowed.   Values of $\kappa_{\cal E}$
within or outside the range, $0\le\kappa_{\cal E}
\le1$, are related to real or imaginary rotation,
respectively.   Values of $\kappa_{\cal E}$ below
or not below unity, are related to bound (finite
extent) and unbound (infinite extent) configurations,
respectively.   Oblate-like $(\epsilon_{31}\le
\epsilon_{21}\le1)$ triaxial configurations lie
between the inner, vertical dotted lines.  Oblate
$(\epsilon_{31}\le\epsilon_{21}=1)$ axisymmetric
configurations lie between the inner and the
corresponding outer,
vertical dotted lines.   Prolate $(\epsilon_{21}
=1\le\epsilon_{31})$ axisymmetric configurations
lie outside the outer, vertical dotted lines.
The equatorial axis ratio is $\epsilon=\epsilon_
{21}$.   The polar axis ratio is $\epsilon=
\epsilon_{31}$ for oblate and $\epsilon=\epsilon_
{13}$ for prolate configurations.   The horizontal
dashed lines define class $E7$ in Hubble (1926)
classification i.e. $0.25<\epsilon\le0.35$ for
edge-on ellipsoids where the line of sight
coincides with the direction of the minor
equatorial axis.}
\label{f:cist}
\end{figure}

The axis ratios of the final configuration,
$\epsilon_{31}$ and $\epsilon_{21}$, as a
function of the parameter, $\kappa_{\cal E}=
{\cal E}_{rot}^\prime/(1-{\cal E}_{osc}^\prime
-{\cal E}_{pec}^\prime)$, are plotted in
Fig.\,\ref{f:cist}, with regard to a Navarro
et al. (1995,1996,1997) density profile $(\Xi=
9.20678$; see CM05, Tab.\,1 therein, for further
details) in rigid rotation.   Each curve is
symmetric with regard to a vertical axis,
$\kappa_{\cal E}=0.5$, where a local minimum
is attained.   
The plane, $({\sf O}\kappa_{\cal E}\epsilon)$,
may be divided into three regions, namely (a)
$0.163190\le\kappa_{\cal E}\le0.836810$, where
triaxial configurations occur, ranging from
$(\epsilon_{21},\epsilon_{31})=(1,0.582724)$ to
$(\epsilon_{21},\epsilon_{31})=(0.363174,0.302297)$,
the latter related to $\kappa_{\cal E}=0.5$; (b)
$0\le\kappa_{\cal E}\le0.163190$, $0.836810\le
\kappa_{\cal E}\le1$, where oblate configurations
occur, ranging from $(\epsilon_{21},\epsilon_{31})=
(1,0.582724)$ to $(\epsilon_{21},\epsilon_{31})=
(1,1)$, the latter related to
$\kappa_{\cal E}=0$ (bound), $\kappa_{\cal E}=1$
(unbound), respectively; (c) $-\infty<\kappa_{\cal
E}\le0$, $1\le\kappa_{\cal E}<+\infty$, where
prolate configurations occur, ranging from
$(\epsilon_{21},\epsilon_{31})=(1,1)$ to
$(\epsilon_{21},\epsilon_{31})=(1,+\infty)$,
the latter related to $\kappa_{\cal E}\to-\infty$
(bound), $\kappa_{\cal E}\to+\infty$ (unbound),
respectively.   It is worth recalling that
oblate-like $(\epsilon_{31}\le\epsilon_{21}\le1)$
and prolate-like $(\epsilon_{21}\le1\le\epsilon_
{31})$ configurations, are related to real and
imaginary rotation, respectively.

An inspection of Fig.\,\ref{f:cist} shows that,
in the case under discussion, cosmological
effects due to expansion prevent dark matter
haloes in real rotation from being more
flattened than $E7$ (dashed horizontal band).
The same holds for embedded giant elliptical
galaxies, provided the related shapes may
safely be thought of as similar and similarly
placed.   On the other hand, an arbitrary
flattening can be attained by dark matter
haloes in imaginary rotation, unless some
kind of instability occurs, which similarly 
prevents configurations more elongated than
$E7$.

As outlined in Subsect.\,\ref{Esel},
bending instabilities have been
suggested long time ago as a viable
mechanism to this respect
(Polyachenko \& Shukhman 1979;
Fridman \& Polyachenko 1984, Vol.\,1,
Chap.\,4, Sect.\,3.3, see also pp.\,313-322;
Vol.\,2, p.\,159; Merritt \& Hernquist 1991;
Merritt \& Sellwood 1994).

The above results depend on both density
profile and rotation velocity profile of
the final configuration, via the coefficient,
$c$, appearing in Eq.\,(\ref{eq:trana}).
Owing to Eqs.\,(\ref{eq:Spq}) and (\ref
{eq:ErJc}), Eq.\,(\ref{eq:tranc}) translates
into:
\begin{leftsubeqnarray}
\slabel{eq:parga}
&& c=\beta_{{\cal E}^\prime}\frac{{\cal R}^\prime}
{({\cal S}^\prime)^2}\frac{(B_{sel})^2}{B_{ram}}
{\cal E}_{rot}\left[\frac{2\zeta-1}{2\zeta}-\frac
{\zeta-1}\zeta{\cal E}_{rot}\right]~~; \\
\slabel{eq:pargb}
&& \beta_{{\cal E}^\prime}=\beta_{\cal E}\frac
{\nu_{sel}^2}{\nu_{ram}}=\frac{\beta_M^5}{\beta_
J^2\beta_E}\frac{\nu_{sel}^2}{\nu_{ram}}~~;
\label{seq:parg}
\end{leftsubeqnarray}
where different changes in mass, angular
momentum, total energy, density profile,
and rotation velocity profile, are related
to the same final shape for an assigned
final configuration, provided $\beta_{{\cal
E}^\prime}$ does not vary.   The choice,
$\beta_{{\cal E}^\prime}=6.007519\cdot10^
{-6}$, in particular $(\beta_M, \beta_J,
\beta_E, \nu_{sel}, \nu_{ram})=(1, 60, 1,
0.610045, 17.20783)$, holds for Fig.\,\ref
{f:cist}.

\section{Conclusion}\label{conc}

With regard to homeoidally striated
Jacobi ellipsoids (CM05), a
unified theory of systematically rotating and
peculiar motions have been developed, where both real
and imaginary rotation (around a fixed axis) have
been considered.   The
effect of residual motion excess along the
equatorial plane or the rotation axis, has been shown
to be equivalent to an additional, real or
imaginary rotation, respectively.   Then it has
been realized that a homeoidally striated Jacobi
ellipsoid with assigned velocity distribution,
always admits an adjoint configuration i.e. a
classical Jacobi ellipsoid of equal mass and
axes.  In addition, further constraints have been
established on the amount of residual velocity
anisotropy along the principal axes, for
triaxial configurations.   Special effort has been
devoted to investigating sequences of virial
equilibrium configurations in terms of
normalized parameters, including the effects
of both density and velocity profile.   In
particular, it has been shown that bifurcation
points from axisymmetric to triaxial 
configurations occur as in classical Jacobi
ellipsoids, contrary to earlier results
(Wiegandt 1982a,b; CM05).
The reasons of the above mentioned discrepancy
have also been explained.   The occurrence of
centrifugal support at the ends of equatorial major
semiaxis, has briefly been discussed.   An
interpretation of recent results from
numerical simulations on stability (Meza
2002), has been provided in the light of the model.
A physical interpretation of the early Hubble
sequence has shortly been reviewed and discussed
from the standpoint of the model.   More
specifically, (i) elliptical galaxies have
been considered as isolated systems, and
an allowed region within
Ellipsoidland (Hunter \& de Zeeuw 1997),
related to the occurrence of bifurcation
points from ellipsoidal to pear-shaped
configurations, has been shown to be consistent
with observations; (ii) elliptical galaxies
have been considered as embedded within dark
matter haloes and, under reasonable
assumptions, it has been shown that tidal effects
from embedding haloes have little influence
on the above mentioned results; (iii) dark
matter haloes and hosted elliptical galaxies,
idealized as a single homeoidally striated
Jacobi ellipsoid, have been considered in connection
with the cosmological transition from expansion to
relaxation, by generalizing an earlier model
(Thuan \& Gott 1975), and the existence of a
lower limit to the flattening of relaxed
(oblate-like) configurations, has been established.
On the other hand, no lower limit has been found 
to the elongation of relaxed (prolate-like)
configurations, and the observed lack of elliptical
galaxies more elongated than $E7$ has needed a
different physical interpretation such as the
occurrence of bending
instabilities (Polyachenko \& Shukhman 1979;
Merritt \& Hernquist 1991).

\section*{Acknowledgements}
We are indebted to D. Merritt and
V. Polyachenko for pointing our attention
to their (and coauthors') papers on bending
instabilities, which had been overlooked
in an earlier version of the current paper.

\appendix
\section{Some properties of ellipsoid
shape factors}
\label{fafo}

Ellipsoid shape factors, $A_1$, $A_2$, $A_3$,
obey the following inequalities (Caimmi 1996a):
\begin{leftsubeqnarray}
\slabel{eq:disAa}
&& a_1^nA_1\ge a_2^nA_2\ge a_3^nA_3~~;\qquad
a_1\ge a_2\ge a_3~~;\qquad n\ge2~~; \\
\slabel{eq:disAb}
&& a_1^nA_1\le a_2^nA_2\le a_3^nA_3~~;\qquad
a_1\ge a_2\ge a_3~~;\qquad n\le1~~;
\label{seq:disA}
\end{leftsubeqnarray}
where inequalities (\ref{eq:disAa}) and
(\ref{eq:disAb}), the latter restricted
to $n\le0$, come from analytical
considerations involving the explicit
expression of the shape factors (e.g.,
MacMillan 1930, Chap.\,II, \S\,33; $A_
1=\alpha^{-2}$, $A_2=\beta^{-2}$, $A_3
=\gamma^{-2}$, therein), and inequality
(\ref{eq:disAb}), restricted to $0<n\le
1$, results from an obvious generalization
of a proof by Pacheco et al. (1989).

Using Eqs.\,(\ref{eq:Bspq}) and (\ref
{eq:Spq}) yields:
\begin{equation}
\label{eq:rsq3}
\frac{{\cal S}_{qq}}{{\cal S}_{33}}=\frac
{A_q}{\epsilon_{3q}^2A_3}~~;\qquad q=1,2~~;
\end{equation}
and, owing to inequality (\ref{eq:disAa}):
\begin{equation}
\label{eq:disAe}
A_q\ge\epsilon_{3q}^2A_3~~;\qquad a_q\ge
a_3~~;\qquad q=1,2~~;
\end{equation}
which implies ${\cal S}_{qq}\ge{\cal S}_{33}$
for oblate-like configurations $(a_q\ge a_3)$,
and ${\cal S}_{qq}\le{\cal S}_{33}$ for
prolate-like configurations $(a_q\le a_3)$.

In the limit of axisymmetric configurations,
$\epsilon_{21}=1$, $\epsilon_{31}=\epsilon$,
$A_1=A_2=\alpha$, $A_3=\gamma$, and the
following relations hold (e.g., Caimmi 1991,
1993):
\begin{lefteqnarray}
\label{eq:lim0}
&& \lim_{\epsilon\to0}\alpha=0~~;\quad
\lim_{\epsilon\to0}\gamma=2~~;\quad\lim_
{\epsilon\to0}\frac\alpha\epsilon=\frac
\pi2~~; \\
\label{eq:limi}
&& \lim_{\epsilon\to+\infty}\alpha=1~~;\quad
\lim_{\epsilon\to+\infty}\gamma=\lim_{\epsilon
\to+\infty}(\epsilon\gamma)=0~~;
\quad\lim_{\epsilon
\to+\infty}(\epsilon^2\gamma)=+\infty~~; \\
\label{eq:liag}
&& \lim_{\epsilon\to1}\frac{\gamma-\alpha}
{1-\epsilon^2}=\frac25~~; \\
\label{eq:dalf}
&& \frac{\diff\alpha}{\diff\epsilon}=\frac1
{1-\epsilon^2}\left[(1+2\epsilon^2)\frac\alpha
\epsilon-2\epsilon\right]~~; \\
\label{eq:dgam}
&& \frac{\diff\gamma}{\diff\epsilon}=\frac1
{1-\epsilon^2}\left[(1+2\epsilon^2)\frac\gamma
\epsilon-\frac2\epsilon\right]~~; \\
\label{eq:upsi}
&& \upsilon=(\upsilon_N)_{iso}=\alpha-
\epsilon^2\gamma~~;
\end{lefteqnarray}
and keeping in mind the general property
(e.g., Chandrasekhar 1969, Chap.\,3,
\S\,17):
\begin{equation}
\label{eq:sA}
A_1+A_2+A_3=2~~;
\end{equation}
it can be seen that the first derivative of
the rotation parameter, $\upsilon$, is null
provided the transcendental equation:
\begin{equation}
\label{eq:maxi}
\alpha=\frac{6\epsilon^2}{1+8\epsilon^2}~~;
\end{equation}
is satisfied.   One solution is found to
exist, which is related to oblate
configurations.   It corresponds to an
extremum point
where the function attains its maximum value
(e.g., Chandrasekhar 1969, Chap.\,5, \S\,32),
as represented in Fig.\,\ref{f:siso}.

\section{Wiegandt criterion for bifurcation}
\label{crib}

Given a collisionless, self-gravitating
fluid in rigid rotation, where no internal
energy transport occurs and the residual
velocity is constant on the boundary, an
upper limit for the point of bifurcation 
is (Wiegandt 1982a,b):
\begin{equation}
\label{eq:bifg}
\Omega^2I_{11}=V_{12;12}~~;
\end{equation}
where $V_{pq;rs}$ is the ``super-matrix'':
\begin{lefteqnarray}
\label{eq:supm}
&& V_{pq;rs}=\int_S\rho(x_1,x_2,x_3)x_p\frac
{\partial{\cal V}_{rs}}{x_q}\diff^3S~~; \\
\label{eq:gpot}
&& {\cal V}_{rs}(x_1,x_2,x_3)=G\int_S\frac
{\rho(x_1^\prime,x_2^\prime,x_3^\prime)(x_p-
x_p^\prime)(x_s-x_s^\prime)}{[(x_1-x_1^\prime)
^2+(x_2-x_2^\prime)^2+(x_3-x_3^\prime)^2]^{3/2}}
\diff^3S~~;
\end{lefteqnarray}
which is expressed in terms of a
generalized potential, ${\cal V}_
{rs}$, and the integrations are
carried over the whole volume, $S$,
of the system.

In the special case of homeoidally
striated Jacobi ellipsoids,
Eq.\,(\ref{eq:bifg}) identifies
the exact point of bifurcation
(Wiegandt 1982a,b) and the
following relations hold (Wiegandt
1982b):
\begin{lefteqnarray}
\label{eq:supq}
&& V_{pq;pq}=-\frac{A_p-a_q^2A_{pq}}{A_p}
(E_{sel})_{pp}~~; \\
\label{eq:Apqr}
&& a_q^2A_{pq}=\frac{A_p-A_q}{1-\epsilon_
{pq}^2}~~;
\end{lefteqnarray}
where the products, $a_q^2A_{pq}$,
are shape factors which depend on
the axis ratios only, similarly to
$A_p$, and symmetry with respect
to the indices holds, $A_{pq}=A_{qp}$.
In addition, Eq.\,(\ref{eq:Apqr})
has been deduced from Chandrasekhar
(1969, Chap.\,3, \S\,21), Eq.\,(107)
therein.

Though Wiegandt (1982b) analysis is
restricted to binomial density
profiles (e.g., Perek 1962;
Chandrasekhar 1969, Chap.\,3, \S\,20;
Caimmi 1993), still it may be
generalized to any kind of cored
density profiles, defined as:
\begin{leftsubeqnarray}
\slabel{eq:depWa}
&& \rho_W(\xi)=\rho_cf(\xi_W)~~;\qquad
f(0)=1~~;\qquad\rho_c=\rho(0)~~; \\
\slabel{eq:depWb}
&& \xi_W=\frac rR~~;\qquad0\le\xi_W\le1
~~;\qquad\Xi_W=1~~;
\label{seq:depW}
\end{leftsubeqnarray}
where the scaling radius and the
scaling density are chosen to be
equal to the radius, $R$, and the
central density, $\rho_c$,
respectively.   The following
relations:
\begin{equation}
\label{eq:codp}
\rho=\frac{\rho_0}{\rho_c}\rho_W~~;
\qquad\xi=\Xi\xi_W~~;
\end{equation}
allow conversion from Eqs.\,(\ref
{seq:profg}) to Eqs.\,(\ref{seq:depW})
and vice versa.

Using Eqs.\,(\ref{seq:depW}), the
mass, the inertia tensor, and the
potential-energy tensor, take the
equivalent expression:
\begin{leftsubeqnarray}
\slabel{eq:MWa}
&& M=(\nu_{mas})_WM_c~~; \\
\slabel{eq:MWb}
&& M_c=\frac{4\pi}3\rho_ca_1^3\epsilon
_{21}\epsilon_{31}~~;
\label{seq:MW}
\end{leftsubeqnarray}
\begin{lefteqnarray}
\label{eq:IpqW}
&& I_{pq}=\delta_{pq}\epsilon_{p1}\epsilon
_{q1}a_1^2M(\nu_{inr})_W~~; \\
\label{eq:EseW}
&& (E_{sel})_{pq}=-(\nu_{sel})_W\frac{GM^2}
{a_1}(B_{sel})_{pq}~~;
\end{lefteqnarray}
and the comparison with Eqs.\,(\ref{eq:M}),
(\ref{eq:Ipq}), (\ref{eq:Espq}), (\ref
{eq:M0}), yields:
\begin{equation}
\label{eq:nuWnu}
(\nu_{mas})_W=\frac{\rho_0}{\rho_c}
\frac1{\Xi^3}\nu_{mas}~~;\qquad
(\nu_{inr})_W=\nu_{inr}~~;\qquad
(\nu_{sel})_W=\nu_{sel}~~;
\end{equation}
finally, using Eqs.\,(\ref{eq:Bspq}) and
(\ref{seq:MW})-(\ref{eq:nuWnu}), the
potential-energy tensor takes the equivalent
form:
\begin{leftsubeqnarray}
\slabel{eq:EsIWa}
&& (E_{sel})_{pq}=-k\pi G\rho_cI_{pq}A_p~~; \\
\slabel{eq:EsIWb}
&& k=\frac43\frac{\rho_0}{\rho_c}\frac
{\nu_{sel}\nu_{mas}}{\Xi^3\nu_{inr}}~~;
\label{seq:EsIW}
\end{leftsubeqnarray}
being $k$, by definition, a profile
factor.

Owing to Eqs.\,(\ref{eq:M0}),
(\ref{eq:hrr}), (\ref{eq:rri}), (\ref
{eq:vg}), (\ref{eq:rome}), (\ref
{eq:EsIWb}), the normalized rotation
parameter, defined by Eq.\,(\ref
{eq:upzN}), may be expressed as:
\begin{equation}
\label{eq:vNW}
\upsilon_N=\frac{\Omega^2}{k\pi G\rho_c}~~;
\end{equation}
which coincides with the parameter,
$\phi$, defined in Wiegandt (1982b),
as shown in Caimmi (1996b).   Accordingly,
Eq.\,(\ref{eq:uNq3}) coincides with its
Wiegandt (1982b) counterpart, Eq.\,(52)
therein.

The combination of  Eqs.\,(\ref{eq:bifg}),
(\ref{eq:supq}), (\ref{seq:EsIW}), (\ref
{eq:vNW}), yields:
\begin{equation}
\label{eq:bivN}
\upsilon_N=A_1-a_2^2A_{12}~~;
\end{equation}
and the comparison with Eq.\,(\ref{eq:uNq3})
reads:
\begin{equation}
\label{eq:biA2}
a_2^2A_{12}=\frac{\zeta_{11}}{\zeta_{33}}
\epsilon_{31}^2A_3~~;
\end{equation}
in the limit of axisymmetric
configurations, $a_2\to a_1$,
$A_{12}\to A_{11}$, the left-hand
side of Eq.\,(\ref{eq:biA2})
takes the expression (e.g., Caimmi
1993):
\begin{equation}
\label{eq:bibi}
\lim_{a_2\to a_1}a_2^2A_{12}=\frac14\left(
3A_1-\epsilon_{31}^2\frac{A_3-A_1}{1-
\epsilon_{31}^2}\right)~~;
\end{equation}
and the combination of Eqs.\,(\ref
{eq:biA2}), (\ref{eq:bibi}), yields:
\begin{equation}
\label{eq:bifW}
\frac{\epsilon_{31}^2A_3}{A_1}=\left[
\frac{5-4\epsilon_{31}^2}{3-2\epsilon_
{31}^2}+\frac{4(1-\epsilon_{31}^2)}
{3-2\epsilon_{31}^2}\frac{\zeta_{11}-
\zeta_{33}}{\zeta_{33}}\right]^{-1}
~~;\qquad\epsilon_{21}=1~~;
\end{equation}
which is an explicit expression of
Wiegandt (1982a,b) criterion for
bifurcation, with regard to homeoidally
striated Jacobi ellipsoids in rigid
rotation.   In the limit of isotropical
residual velocity distribution,
$\zeta_{11}=\zeta_{22}=\zeta_{33}$,
Eq.\,(\ref{eq:bifW}) reduces to (\ref
{eq:bifJ}), as expected.

The condition, defined by Eq.\,(\ref
{eq:bifW}), has to be compared with
its counterpart deduced in the current
paper, under the same assumption i.e.
$\zeta_{11}=\zeta_{22}$ also for
triaxial configurations.   The latter
may be deduced from Eq.\,(\ref{eq:bifz}),
using the relation (e.g., Caimmi 1966a):
\begin{equation}
\label{eq:bifc}
\lim_{\epsilon_{21}\to1}\epsilon_{21}^2
\frac{A_2-A_1}{1-\epsilon_{21}^2}=\frac
14\left(3A_1-\epsilon_{31}^2\frac{A_3-
A_1}{1-\epsilon_{31}^2}\right)~~;
\end{equation}
and the result is again Eq.\,(\ref
{eq:bifW}).   Then Wiegandt (1982b)
criterion for bifurcation can be
deduced from the current theory,
provided isotropic residual velocity
distribution is assumed along the
equatorial plane, also for triaxial
configurations.   On the other hand,
the above assumption is in contradiction
with Eqs.\,(\ref{seq:z12}) and then
it cannot be accepted.   A criterion
for bifurcation, consistent with
Eqs.\,(\ref{seq:z12}), is expressed
by Eq.\,(\ref{eq:bifJ}).

The origin of the discrepancy could be
the following.   Wiegandt (1982a)
assumption (ii) implies Eq.\,(12)
therein which, after integration,
yields isotropic residual velocity
distribution along the equatorial
plane i.e. $\zeta_{11}=\zeta_
{22}$.   On the other hand, an isotropic
residual velocity distribution along
the equatorial plane is related to a
zero-th order approximation in some
special series developments [Marochnik
1967; Eq.\,(19) therein].   The additional
requirement of axial symmetry makes the
above mentioned relation be exact, but in
presence of triplanar symmetry it has to
be considered as a zero-th order
approximation.

In the light of the results discussed in
the current subsection, the
analogy between the behaviour of
collisional and collisionless
self-gravitating fluids, subjected
to the restrictions (Wiegandt 1982a,b):
(i) rigid rotation; (ii) constant
residual velocity on the boundary;
(iii) absence of internal energy
transports; appears to be complete.


\begin{thebibliography}{}
\bibitem{} Bertola, F., Capaccioli, M. 1975, ApJ 200, 439
\bibitem{} Bertola, F., Galletta, G. 1978, ApJ 226, L115
\bibitem{} Bett, P. et al. 2007, MNRAS 376, 215
\bibitem{} Binney, J. 1976, MNRAS 177, 19
\bibitem{} Binney, J. 1978, MNRAS 183, 501
\bibitem{} Binney, J. 1980, MNRAS 190, 421
\bibitem{} Binney, J., \& Tremaine, S. 1987, {\it Galactic
           Dynamics}, Princeton University Press, Princeton
\bibitem{} Blaauw, A. 1965, in {\it Stars and Stellar Systems},
           vol.\,V, Chap.\,20, \S\,3     
\bibitem{} Brosche, P., Caimmi, R., \& Secco, L. 1983, A\&A 125, 338
\bibitem{} Caimmi, R. 1979, ApSS 71, 75
\bibitem{} Caimmi, R. 1983, ApSS 93, 403
\bibitem{} Caimmi, R., 1991, ApSS 180, 211
\bibitem{} Caimmi, R. 1993a, ApJ 419, 615 
\bibitem{} Caimmi, R. 1993b, ApSS 199, 11
\bibitem{} Caimmi, R. 1996a, Acta Cosmologica XXII, 21
\bibitem{} Caimmi, R. 1996b, AN 317, 401
\bibitem{} Caimmi, R. 2003, AN 324, 250
\bibitem{} Caimmi, R., Secco, L., Brosche, P. 1984, A\&A 139, 411  
\bibitem{} Caimmi, R., Secco, L. 1992, ApJ 395, 119  
   \bibitem{} Caimmi, R., \& Marmo, C. 2003, NewA 8, 119 (CM03) 
   \bibitem{} Caimmi, R., \& Marmo, C. 2005, AN, submitted (CM05)
\bibitem{} Chandrasekhar, S., \& Leboviz, N.R. 1962, ApJ 136, 1082
\bibitem{} Chandrasekhar, S. 1969, {\it Ellipsoidal Figures
             of Equilibrium}, Yale University Press, New Haven
\bibitem{} Durisen, R.H. 1978, ApJ 224, 826
\bibitem{} Fridman, A., Polyachenko, V.L. 1984, {\it Physics of Gravitating
           Systems}, Springer-Verlag
\bibitem{} Hernquist, L. 1990, ApJ 356, 359
\bibitem{} Hoeft, M., M\"ucket, J.P., Gottl\"ober, S. 2004,
           ApJ 602, 162
\bibitem{} Hubble, E. 1926, ApJ 64, 321
\bibitem{} Hunter, C., \& de Zeeuw, T.Z. 1997, ApJ 389, 79
\bibitem{} Illingworth, G. 1977, ApJ 218, L43
\bibitem{} Illingworth, G. 1981, in S. M. Fall and D. Lynden-Bell (eds.),
           {\it Structure and Evolution of Normal Galaxies}, Cambridge
           University Press, p.\,27
\bibitem{} Jeans, J. 1929, {\it Astronomy and Cosmogony},
             Dover Publications, New York, 1961
\bibitem{} Lai, D., Rasio, F.A., \& Shapiro, S.L. 1993, ApJS 88, 205 
\bibitem{} Landau, L., Lifchitz, E. 1966, {\it Mecanique}, Mir, Moscow
\bibitem{} MacMillan, W.D. 1930, {\it The Theory of the Potential},
           Dober Publications, New York, 1958
\bibitem{} Marochnik, L.S. 1967, Soviet Astron. AJ 10, 738
\bibitem{} Meza, A. 2002, A\&A 395, 25
\bibitem{} Merritt, D., Aguilar, L.A. 1985, MNRAS 217, 787 
\bibitem{} Merritt, D., Hernquist, L. 1991, ApJ 376, 439
\bibitem{} Merritt, D., Sellwood, J. 1994, ApJ 425, 551
\bibitem{} Navarro, J.F., Frenk, C.S., \& White, S.D.M. 1995,
            MNRAS 275, 720
\bibitem{} Navarro, J.F., Frenk, C.S., \& White, S.D.M. 1996, 
            ApJ 462, 563 
\bibitem{} Navarro, J.F., Frenk, C.S., \& White, S.D.M. 1997, 
            ApJ 490, 493
\bibitem{} Nipoti, C., Londrillo, P., \& Ciotti, L. 2002, MNRAS 332, 901
\bibitem{} Pacheco, F., Pucacco, G., \& Ruffini, R. 1986, A\&A 161, 39
\bibitem{} Pacheco, F., Pucacco, G., Ruffini, R., \& Sebastiani, G. 1989, 
           A\&A 210, 42
\bibitem{} Perek, L. 1962, Adv. Astron. Astrophys. 1, 165
\bibitem{} Polyachenko, V.L., Shukhman, I.G. 1979, Sov. Astr. 27, 407
\bibitem{} Raha, N., et al. 1991, Nature 352, 411
\bibitem{} Rasia, E., Tormen, G., Moscardini, L. 2004, MNRAS 351, 237
\bibitem{} Roberts, P.H. 1962, ApJ 136, 1108
\bibitem{} Roberts, M.S., Haynes, M.P. 1994, ARAA 32, 115
\bibitem{} Schechter, P.L., Gunn, J.E. 1979, ApJ 229, 472
\bibitem{} Stiavelli, M., Sparke, L.S. 1991, ApJ 382, 466
\bibitem{} Thuan, T.X., \& Gott, J.R., III, 1975, Nature 257, 774
\bibitem{} van den Bergh, S. 1960a, ApJ 131, 215
\bibitem{} van den Bergh, S. 1960b, ApJ 131, 558
\bibitem{} Vandervoort, P.O., 1980a, ApJ 240, 478
\bibitem{} Vandervoort, P.O., 1980b, ApJ 241, 316
\bibitem{} Vandervoort, P.O., Welty, D.E. 1981, ApJ 248, 504
\bibitem{} Wiegandt, R., 1982a, A\&A 105, 326
\bibitem{} Wiegandt, R., 1982b, A\&A 106, 240

\end{thebibliography}
\end{document}